\documentclass[prd,aps,
preprint,
superscriptaddress,showpacs,eqsecnum
,nofootinbib
,amsmath,amssymb
]{revtex4}

\usepackage{graphics,graphicx,epsfig,color
}

\usepackage[T1]{fontenc} 

\def\pbar{\overline{p}}
\def\be{\begin{equation}}
\def\ee{\end{equation}}

\def\gsim{\mathrel{
\rlap{\raise 0.511ex \hbox{$>$}}{\lower 0.511ex
\hbox{$\sim$}}}}
\def\lsim{\mathrel{
\rlap{\raise 0.511ex \hbox{$<$}}{\lower 0.511ex
\hbox{$\sim$}}}}
\newcommand{\G}{\overline {g}}

\usepackage{ifpdf}
\ifpdf
\usepackage[colorlinks=true, pdfstartview=FitV, linkcolor=red, 
citecolor=blue, urlcolor=magenta, breaklinks=true]{hyperref}
\fi
\begin{document}
\title{
A geometric algorithm for efficient coincident detection \\ of gravitational waves}
\author{C.A.K. Robinson}
\email{C.Robinson@astro.cf.ac.uk}

\author{B.S. Sathyaprakash}
\email{B.Sathyaprakash@astro.cf.ac.uk}
\affiliation{School of Physics and Astronomy, Cardiff University, 5, The Parade, Cardiff, UK, CF24 3AA}

\author{Anand S. Sengupta}
\email{sengupta@ligo.caltech.edu}
\affiliation{School of Physics and Astronomy, Cardiff University, 5, The Parade, Cardiff, UK, CF24 3AA}
\affiliation{LIGO Laboratory, California Institute of Technology, Pasadena CA 91125, USA}

\begin{abstract}
Data from a network of gravitational wave detectors can be
analyzed in coincidence to increase detection 
confidence and reduce non-stationarity of
the background. We propose and explore a geometric 
algorithm to combine the data from a network of detectors. 
The algorithm makes optimal use of the variances and 
covariances that exist amongst the different parameters 
of a signal in a coincident detection of events. The new 
algorithm essentially associates with each trigger ellipsoidal 
regions in parameter space defined by the covariance matrix. Triggers from 
different detectors are deemed to be in coincidence if their
ellipsoids have a non-zero overlap. Compared
to an algorithm that uses uncorrelated windows separately
for each of the signal parameters, the new algorithm greatly
reduces the background rate thereby increasing detection
efficiency at a given false alarm rate.
\end{abstract}

\preprint{LIGO P080049-00-Z}
\maketitle

\section{Motivation}
Long baseline interferometric gravitational wave detectors,
such as the Laser Interferometer Gravitational-Wave Observatory 
(LIGO) \cite{Abbott:2003vs}, Virgo \cite{Acernese:2006bj}, and 
GEO 600 \cite{Luck:2006ug}, are currently acquiring the best data 
ever. The data sets from the different detectors can be either brought
together and analyzed phase coherently \cite{Pai:2000zt,Bose:1999bp,Finn:2000hj,Arnaud:2003zq}, 
or analyzed separately followed up by a coincidence analysis
\cite{Jaranowski:1994xd,Jaranowski:1994,Finn:2000hj,Arnaud:2001my,Tagoshi:2007ni,Abbott:2003pj,Abbott:2005pe,Abbott:2005pf,Abbott:2005kq} 
of the triggers obtained. 
Coherent analysis maximizes signal visibility (i.e., gives the best 
possible signal-to-noise ratio in the likelihood sense) while the goal 
of coincidence analysis is to reduce and mitigate the non-stationary 
and non-Gaussian background noise. A recent comparison of coherent 
analysis vis-a-vis coincidence analysis under the assumption that 
the background noise is Gaussian and stationary has concluded that 
coherent analysis, as one might expect, is far better than coincidence 
analysis~\cite{CoherentvsCoincident}. However, there are two reasons why current data analysis 
pipelines prefer the latter over former. Firstly, since the detector 
noise is neither Gaussian nor stationary, coincidence analysis can 
potentially reduce the background rate far greater than one might 
think otherwise. Secondly, coherent analysis is computationally far 
more expensive than coincidence analysis and it is presently not 
practicable to employ coherent analysis.

\subsection{The problem of coincident detection}

In coincidence analysis 
(see for example, Refs.\ 
\cite{Abbott:2003pj,Abbott:2005fb,Abbott:2005pe,Abbott:2005pf,Abbott:2005at,Abbott:2005kq}), 
data sets from each detector will
be analyzed separately and the triggers from the end of the pipeline
from different detectors compared with one another to identify triggers 
that might be in coincidence with one another. More precisely, the goal 
is to find if the parameters of a trigger (e.g., in the case of a coalescing binary the 
time of merger, the component masses and spins) from one detector 
are identical to those from another. Since the presence of noise causes
errors in the measurement of parameters of an
inherent signal, it is highly improbable that the same gravitational wave
in different detectors can be associated with exactly the same set of
parameters. However, it should be possible to detect signals in coincidence
by demanding that the measured parameters lie in a sufficiently small range of each other
\cite{Abbott:2003pj,Abbott:2005fb,Abbott:2005pe,Abbott:2005pf,Abbott:2005at,Abbott:2005kq}. 
Thus, we can revise the coincidence criteria as follows:
{\em triggers from different detectors are said to be in coincidence if 
their parameters all lie within a certain acceptable range}. Events that pass 
the coincidence test are subject to further scrutiny but we shall focus 
in this paper on the coincidence test itself.

From the above discussion it is clear that an important 
aspect of coincidence analysis is the determination of the range of
parameter values to be associated with each trigger. To this end, until 
recently, the LIGO Scientific Collaboration (LSC) has deployed a 
phenomenological method for assigning the ranges \cite{Abbott:2003pj,Abbott:2005fb,Abbott:2005pe,Abbott:2005pf,Abbott:2005at,Abbott:2005kq}. 
More precisely,
one performs a large number of simulations in which a signal with a
known set of parameters is added in software to the data which is then
passed through the analysis pipeline. The pipeline identifies
the most probable parameters with each injected signal and the ensemble of
injected and measured parameters gives the distribution of the errors
incurred in the measurement process. Given the distribution of the 
errors, one can choose a range for each parameter such that more than,
say, 95\% of the injected signals are detected in coincidence. Choosing
wider windows will enable greater detection probability but also increases
the rate of accidental triggers. On the contrary, smaller windows decrease
the false alarm rate but also reduce the detection probability.
Recently, a Bayesian coincidence test has been proposed \cite{Cannon:2008}
as an alternative wherein one computes the likelihood of a 
candidate event as  belonging to a distribution obeyed by true
signals rather the noise background. Unfortunately, measuring the
distribution function when the parameter space is large could be 
computationally formidable \cite{Cannon:2008} except when the 
parameters are independent.

\subsection{A geometric approach to choosing coincident windows}

In this paper we propose a new algorithm based on the metric 
(equivalently, the information matrix) defined on the signal manifold.
The idea is very simple, even obvious, but leads to a great reduction
in the background trigger rate. The advantages of the new algorithm are
better appreciated by listing certain drawbacks of the phenomenological 
method. The drawbacks are quite naturally remedied in the new approach. 

First, because the current method uses rectangular windows 
it ignores the correlations between different parameters. For instance,
in the case of a chirping signal from a black hole binary the shape of
the signal depends, among others, on the component masses. However, not all 
combinations of the two masses lead to signals that are easily distinguishable 
from one another. Indeed, at the lowest post-Newtonian order the
waveform depends only on a certain combination of the masses called the
{\em chirp mass}; binaries of different values for the two masses but the
same chirp mass produce essentially the same signal. This degeneracy
is broken when post-Newtonian corrections are included. Nevertheless, 
the two mass parameters continue to be highly correlated. 

The second drawback is that the method employs windows of the same size 
throughout the parameter space while we know that errors in the measurement
of the parameters depends, in some cases quite sensitively, on the
parameters. Drawing again from our example of a binary, the error in
the estimation of the chirp mass can vary by more than two orders of
magnitude across the parameter space of interest in the case of systems
that LIGO is expected to observe (see, e.g., 
\cite{CF94,Finn92,FinnCh93,Chernoff:1993th,KKT94,KKS95,PW95,BalSatDhu95,BalSatDhu96,FlanHugh97,AISS05}). 
Clearly, it is not optimal to deploy
windows of the same size all over the parameter space. 

Thirdly, by not taking into account parameter covariances, the method 
entails independent tuning of several parameters at the same time. 
This could be a horrendous problem when dealing with signals characterized
by many parameters. For instance, continuous radiation from a pulsar
is characterized by the location of the pulsar, its spin frequency,
the derivative of the frequency and so on. These physical parameters
are all not independent; the existence of covariances among them imply 
that the effect of variation in one parameter can be absorbed by another - 
thereby complicating the pipeline tuning procedure. In the case where parameters 
have perfect or near perfect covariances, variations of the parameters may not 
even lead to distinct signals at all. This further implies that it may not be necessary to 
tune each parameter separately, rather it should be enough to tune only a subset of the
parameters or, more precisely, only the principal components. Furthermore,
the method does not provide a unique set of windows, rather several
possibilities could be worked out.

Finally, by using windows of the same size irrespective of the
signal-to-noise ratio (SNR) of the trigger, the method suffers from an
undesirably high false alarm rate, particularly in the tail of the SNR distribution.
Needless to say,  a successful detection of gravitational waves necessitates
as clean a distribution of the SNRs as possible, with little contamination of
the tails. One way of reducing the false alarm rate is by using tighter 
windows at higher SNRs. This is well-motivated since true high-SNR events
will be associated with smaller errors.

The geometric algorithm proposed in this paper quite naturally overcomes 
the drawbacks  of the phenomenological method. 
The algorithm takes into account the correlations amongst 
the various parameters and deploys parameter- and SNR-dependent ellipsoidal 
windows defined by the Fisher information matrix using a single parameter. 
The most important consequence of the new algorithm is a great 
reduction in the background rate.

\subsection{Organization of the paper}
In Secs.\ \ref{sec2},\ \ref{subsec:ellipsoidOverlap} and\ \ref{sec:reduction in FAR},
we present and discuss the new algorithm to identify
events in coincidence. The algorithm comprises of two steps. The first step 
consists in associating each trigger with a $p$-dimensional ellipsoid. 
In the second step one tests 
if the ellipsoid associated with a trigger from one detector 
overlaps, or at least touches, an ellipsoid associated with a trigger
from another detector. 
In Sec.\ \ref{sec3} we apply the algorithm
developed in Sec.\ \ref{sec2} to the case of a  
transient chirp signal from
a binary black hole. This will help us assess the extent to
which the algorithm is helpful in reducing the background. Sec.\
\ref{sec4} concludes by summarizing the application of the 
new algorithm in real data analysis pipelines and future prospects.

\section{A geometric coincidence algorithm}
\label{sec2}
This Section begins with a brief introduction to the geometric
formulation of signal manifold and metric introducing the terminology 
needed in later Sections. The metric so defined helps us in identifying
ellipsoidal regions with a given point on the manifold whose size is
chosen depending on the signal-to-noise ratio (SNR) and the parameter space
region where the point lies. As an exercise to estimate the efficacy of the new coincidence algorithm we then compare the volume of the ellipsoid
with that of a proper rectangular box enclosing the ellipsoid and
aligned along the coordinate lines.
 
\subsection{Scalar Product, Signal Manifold and Metric}
\label{sec2:ScalProd}
The problem of gravitational wave data analysis was addressed
in a geometric framework with the intention of understanding
parameter estimation \cite{BalSatDhu95,BalSatDhu96} and computational 
requirements for matched filtering \cite{SathyaFilter94,Owen96,OwenSathyaprakash98}. 
In this framework, one thinks of the outputs of an ensemble 
of detectors as either finite- or infinite-dimensional vectors 
depending on whether one considers data streams as a
discrete sampled set or the continuum limit of the same,
respectively. For the sake of convenience, in this paper 
we shall deal with the continuum limit. However, all our
results are applicable to the more realistic case in which
detector outputs are treated as finite dimensional vectors.
It is easy to see that the set of all detector outputs 
form a vector space satisfying the usual axioms of a vector space.
The starting point of our discussion is the definition of
the scalar product. Given any two functions $x(t)$ and $y(t),$ 
their scalar product $\left<x,\,y\right>$ is defined as 
\cite{Finn92,FinnCh93,Chernoff:1993th,CF94}
\begin{equation}
\left< x,\, y\right >  = 2 \int_0^\infty \frac{{\rm d}f}{S_h(f)} 
\left [ X(f)\, Y^*(f) + X^*(f) Y(f) \right ],
\label{eq:sp}
\end{equation}
where $X(f) \equiv \int^\infty_{-\infty} {\rm d}t\, x(t)\, \exp(-2\pi i f t)$
is the Fourier transform of the function $x(t)$ (and similarly, $Y(f)$) and
$S_h(f)$ is the one-sided noise power-spectral density of the detector.
The scalar product in Eq.~(\ref{eq:sp}) is motivated by the 
likelihood of a known signal buried in Gaussian, stationary 
background \cite{Helstrom68}. 

Amongst all vectors, of particular interest are those corresponding
to gravitational waves from a given astronomical source. While every signal
can be thought of as a vector in the infinite-dimensional vector space of
the detector outputs, the set of all such signal vectors don't, by themselves, 
form a vector space.
One can immediately see that the norm of a signal $h$ (i.e., the square-root of
the scalar product of a signal with itself) gives the SNR
$\rho$ for a noiseless signal that is filtered using an optimal template \cite{Th300,Schutz89}:
\begin{equation}
\rho \equiv \left< h,\, h\right >^{1/2}  = 2 \left [ \int_0^\infty 
\frac{{\rm d}f}{S_h(f)} \left | H(f)\right |^2 \right ]^{1/2},
\label{eq:norm}
\end{equation}
where $H(f)$ is the Fourier transform of the signal $h(t).$
In particular, we can define signals $\hat h$ of unit norm:
\begin{equation}
{\hat h} \equiv \frac{h}{\sqrt{\left<h,\, h\right >}}
= \frac{h}{\rho},\ \ \ \ 
\left<\hat h,\, \hat h\right >=1.
\label{eq:unitnorm}
\end{equation}
 
The set of all normed signal vectors (i.e., signal vectors of unit norm)
form a manifold, the parameters of the signal serving as a coordinate system 
\cite{BalSatDhu95,BalSatDhu96,Owen96,OwenSathyaprakash98}. 
Thus, each class of astronomical source 
forms an $n$-dimensional manifold ${\cal S}_n,$ where $n$ is the number of 
independent parameters characterizing the source. For instance, the 
set of all signals from a binary on a quasi-circular orbit inclined
to the line-of-sight at an angle $\iota,$ consisting of non-spinning 
black holes of masses $m_1,$ and $m_2,$ located a distance $D$ from 
the Earth\footnote{Even though we deal with normed signals (which amounts to fixing $D$), astrophysical gravitational wave signals are characterised by this additional parameter.} initially in the direction $(\theta,\varphi)$ and expected 
to merge at a time $t_C$ with the phase of the signal at merger 
$\varphi_C$, forms a nine-dimensional manifold with coordinates 
$\{D,\,\theta,\, \varphi,\, m_1,\, m_2,\, t_C,\, \varphi_C,\,
\iota,\,\psi\},$ where $\psi$ is the polarization angle of the signal.
In the general case of a signal characterized by $n$ parameters we 
shall denote the parameters by $p^{\alpha},$ where $\alpha=1,\ldots,n.$

The manifold ${\cal S}_n$ can be endowed with a metric $g_{\alpha\beta}$ 
that is induced by the scalar product defined in Eq.~(\ref{eq:sp}). 
The components of the metric in a coordinate system $p^\alpha$ 
are defined by\footnote{We have followed the definition of the metric
as is conventional in parameter estimation theory (see, e.g., Refs.\ 
\cite{Finn92,FinnCh93,Chernoff:1993th,BalSatDhu96})
which differs from that used in template placement algorithms (see, e.g.
Refs.\ \cite{Owen96}) by a factor of 2. This difference will impact
the relationship between the metric and the match as will 
be apparent in what follows.}
\begin{equation}
g_{\alpha\beta} \equiv 
\left < \partial_\alpha \hat h,\, \partial_\beta \hat h \right >,
\ \ \ \ \partial_\alpha \hat h \equiv \frac{\partial \hat h}{\partial p^\alpha}.
\label{eq:metricDef}
\end{equation}
The metric can then be used on the signal manifold as a measure of the proper 
distance ${\rm d}\ell$ between nearby signals with coordinates $p^\alpha$ and 
$p^\alpha+{\rm d}p^\alpha,$ that is signals $\hat h(p^\alpha)$ and 
$\hat h(p^\alpha+{\rm d}p^\alpha)$, 
\begin{equation}
\label{eq:properDistance}
{\rm d}\ell^2 = 	g_{\alpha\beta} {\rm d}p^\alpha {\rm d}p^\beta.
\end{equation}

Now, by Taylor expanding $\hat h(p^\alpha+{\rm d}p^\alpha)$ around $p^\alpha,$
and keeping only terms to second order in ${\rm d}p^\alpha,$
it is straightforward to see that the overlap $\cal O$ of two infinitesimally 
close-by signals can be computed using the metric:
\begin{eqnarray}
	\label{eq:overlap}
	{\cal O}({\rm d}p^\alpha;\, p^\alpha) & \equiv & \left <\hat h(p^\alpha),\, 
	\hat h(p^\alpha+{\rm d}p^\alpha) \right >\nonumber \\
	         & = & 1 - \tfrac{1}{2} g_{\alpha\beta} {\rm d}p^\alpha {\rm d}p^\beta,
\end{eqnarray}

The metric on the signal manifold is nothing but the well-known
Fisher information matrix usually denoted $\Gamma_{\alpha\beta}$,
(see, e.g., \cite{Helstrom68, pBCV})
but scaled down by the square of the SNR, i.e.,
$g_{\alpha\beta} = \rho^{-2}\Gamma_{\alpha\beta}.$
The information matrix is itself the inverse of the covariance 
matrix $C_{\alpha\beta}$ and is a very useful quantity in signal analysis. 
The ambiguity function ${\cal A}({\rm d}p^\alpha;\, p^\alpha)$, familiar 
to signal analysts, is the overlap function defined above: 
${\cal A}({\rm d}p^\alpha;\, p^\alpha) \simeq {\cal O}({\rm d}p^\alpha;\, p^\alpha).$ 
Thus, the equation
\begin{equation}
\label{eq:ambiguity}
{\cal A}({\rm d}p^\alpha;\, p^\alpha) = \epsilon,\ \ {\rm or}\ \ 
{\cal O}({\rm d}p^\alpha;\, p^\alpha) = \epsilon,
\end{equation} 
where $\epsilon$ ($0< \epsilon < 1$) is a constant, defines the ambiguity surface, 
or level surface. In gravitational wave literature $\epsilon,$ which 
measures the overlap between two mis-matched signals, is also called the 
{\em match}. Using the expression for the overlap ${\cal O}$
[cf.\ Eq.~(\ref{eq:overlap})] in Eq.~(\ref{eq:ambiguity}), we can see that
the coordinate distance ${\rm d}p^{\alpha}$ to the ambiguity surface from
the coordinate point $p^\alpha$ is related to the proper distance\footnote{Here
the proper distance refers to the distance between the signal $\hat h(p^\alpha)$ 
at the coordinate point $p^\alpha$ and a signal $\hat h(p^\alpha+{\rm d}p^\alpha)$ 
with coordinates $p^\alpha+{\rm d}p^\alpha$ on the ambiguity surface.} by:
\begin{equation}
	\label{eq:ellipsoid}
	g_{\alpha\beta} {\rm d}p^\alpha {\rm d}p^\beta = 2(1 - \epsilon).
\end{equation}
Equivalently, ${\rm d}\ell = \sqrt{2(1 - \epsilon)}.$
For a given value of the match $\epsilon$ the above equation defines a 
$(n-1)$-dimensional ellipsoid in the  
$n$-dimensional signal manifold. Every signal with parameters 
$p^\alpha+{\rm d}p^\alpha$ on the ellipsoid has an overlap $\epsilon$ 
with the reference signal at $p^\alpha.$

\subsection{Coincidence windows}
Having defined the metric (equivalently, the information matrix)
and the ambiguity function, we next consider the
application of the geometric formalism in the estimation of statistical
errors involved in the measurement of the parameters and then discuss how
that information may be used in coincidence analysis. We closely
follow the notation of Finn and Chernoff 
\cite{Finn92,FinnCh93,Chernoff:1993th} to introduce the basic ideas
and apply their results in the choice of coincidence windows.

Let us suppose a
signal of known shape with parameters $p^\alpha$ is buried in background
noise that is Gaussian and stationary. Since the signal shape is known
one can use matched filtering to dig the signal out of noise. The measured
parameters $\pbar^\alpha$ will, in general, differ from the true parameters 
of the signal\footnote{In what follows we shall use an 
over-line to distinguish the measured parameters from true parameters $p^\alpha$.}. 
Geometrically speaking, the noise vector displaces the signal vector 
and the process of matched filtering projects the (noise + signal) 
vector back on to the signal manifold. Thus, any non-zero noise will make
it impossible to measure the true parameters of the signal. The best one
can hope for is a proper statistical estimation of the influence of noise.

The posterior probability density function ${\cal P}$ of 
the parameters $\pbar^\alpha$ is given by a multi-variate Gaussian distribution
\footnote{A Bayesian interpretation of ${\cal P}(\Delta p^\alpha)$ is the probability of having the true signal parameters to lie somewhere inside the ellipsoidal volume centered at the Maximum Likelihood point $\pbar^\alpha$. In this case the overlap-test for determining coincidence test is motivated by a test of concordance that the true signal parameters $p^\alpha$ should lie in the overlap region.}:
\begin{equation}
\label{eq:multivariatePDF}
{\cal	P}(\Delta p^\alpha)\, {\rm d}^n\Delta p = 
\frac{{\rm d}^n\Delta p}{(2\pi)^{n/2}\sqrt{C}} 
\exp \left [-\frac{1}{2}C^{-1}_{\alpha\beta}\, \Delta p^\alpha \Delta \, p^\beta \right ],
\end{equation}
where $n$ is the number of parameters, $\Delta p^\alpha = p^\alpha  - \pbar^\alpha,$ and 
$C_{\alpha\beta}$ is the covariance matrix, $C$ being its determinant. Noting 
that $C^{-1}_{\alpha\beta}=\rho^2 g_{\alpha\beta},$ we can re-write the above distribution as:
\begin{equation}
\label{eq:multivariatePDFinMetric}
{\cal	P}(\Delta p^\alpha)\, {\rm d}^n\Delta p = 
\frac{\rho^n\, \sqrt{g}\, {\rm d}^n\Delta p}{(2\pi)^{n/2}} 
\exp \left [-\frac{\rho^2}{2}\, g_{\alpha\beta}\, 
\Delta p^\alpha \Delta \, p^\beta \right ].
\end{equation}
where we have used the fact that $C=1/(\rho^{2n}\, g),$ $g$ being the determinant 
of the metric $g_{\alpha\beta}$.  Note that if we
define new parameters $p'^\alpha = \rho p^\alpha,$ then we have exactly the {\em same}
distribution function for all SNRs, except the deviations $\Delta p^\alpha$ are scaled by
$\rho.$ 

Let us first specialize to one-dimension to illustrate what region of the parameter
space one should associate with a given trigger. In one-dimension the distribution
of the deviation from the mean of the measured value of the parameter $p$ is given by:
\begin{equation}
\label{eq:prob}
{\cal P}(\Delta p) {\rm d}\Delta p 
= \frac{{\rm d}\Delta p}{\sqrt{2\pi}\sigma} \exp\left (-\frac{\Delta p^2}{2\sigma^2}\right )
= \frac{\rho\,\sqrt{g_{pp}} {\rm d}\Delta p}{\sqrt{2\pi} } 
\exp\left (-\frac{\rho^2}{2}{g_{pp}\Delta p^2} \right ),
\end{equation}
where, analogous to the $n$-dimensional case, we have used $\sigma^2=1/(\rho^2g_{pp}).$
Now, at a given SNR, what is the volume $V_P$ in the parameter space 
such that the probability of finding the measured parameters $\pbar$ inside this 
volume is $P?$ This volume is defined by:
\begin{equation}
P = \int_{\Delta p \in V_P}{\cal P}(\Delta p) {\rm d}\Delta p.
\label{eq:volumeInsideP}
\end{equation}
Although $V_P$ is not unique it is customary to choose it to be centered around $\Delta p=0:$
\begin{equation}
P = \int_{(\Delta p/\sigma )^2 \le r^2(P)} \frac{{\rm d}\Delta p}{\sqrt{2\pi} \sigma} 
\exp\left (-\frac{\Delta p^2}{2\sigma^2} \right ) 
= \int_{\rho^2 g_{pp}\Delta p^2 \le r^2(P)}\frac{\rho\, \sqrt{g_{pp}} {\rm d}\Delta p}{\sqrt{2\pi}} 
\exp\left (-\frac{\rho^2\,g_{pp}\Delta p^2}{2} \right ),
\end{equation}
where given $P$ the above equation can be used to solve for $r(P)$ and it
determines the range of integration. 
For instance, the volumes $V_P$ corresponding to $P\simeq 0.683, 0.954, 0.997,\ldots,$ are the 
familiar intervals $[-\sigma,\, \sigma],$ $[-2\sigma,\, 2\sigma],$ $[-3\sigma,\, 3\sigma],$ $\ldots,$
and the corresponding values of $r$ are $1,$ $2,$ $3.$
Since $\sigma=1/\sqrt{\rho^2 g_{pp}}$ we see that in terms of $g_{pp}$ the above intervals
translate to 
\begin{equation}
\frac{1}{\rho} \left[-\frac{1}{\sqrt{g_{pp}}},\, \frac{1}{\sqrt{g_{pp}}}\right ],\,\,
\frac{1}{\rho} \left[-\frac{2}{\sqrt{g_{pp}}},\, \frac{2}{\sqrt{g_{pp}}}\right ],\,\,
\frac{1}{\rho} \left[-\frac{3}{\sqrt{g_{pp}}},\, \frac{3}{\sqrt{g_{pp}}}\right ],\ldots.
\end{equation}
Thus, for a given probability $P$, the volume $V_P$ shrinks as $1/\rho.$ The maximum
distance $d_{\rm max}$ within which we can expect to find ``triggers'' 
at a given $P$ depends inversely on the SNR $\rho:$
$d{\ell} = \sqrt{g_{pp}\Delta p^2} = r/\rho.$  Therefore, for $P\simeq 0.954,$ 
$r=2$ and at an SNR of $5$ the maximum distance is $0.4,$ which
corresponds to a match of $\epsilon=1- \tfrac{1}{2}d{\ell}^2
= 0.92.$ In other words, in one-dimension 95\% of the time
we expect our triggers to come from templates that have an overlap  
greater than or equal to 0.92 with the buried signal when the SNR is 5.
This interpretation in terms of the match is a good approximation as long as
$d{\ell} \ll 1,$ which will be true for large SNR events. However, for
weaker signals and/or greater values of $P$ we can't interpret the results
in terms of the match although, the foregoing equation Eq.~(\ref{eq:volumeInsideP}) can be
used to determine $r(P).$ As an example, at $P\simeq 0.997,$ $r=3$ and at an
SNR of $\rho=4$ the maximum distance is $d\ell=0.75$
and the match is $\epsilon=23/32\simeq 0.72,$ which is significantly smaller than 1
and the quadratic approximation is not good enough to compute the match.

These results generalize to $n$ dimensions. In $n$-dimensions the volume $V_P$ 
is defined by
\begin{equation}
\label{eq:VolumeDefiningEqn}
P = \int_{\Delta p^\alpha \in V_P} {\cal P}(\Delta p^\alpha)\, 
{\rm d}^n\Delta p. 
\end{equation}
Again, $V_P$ is not unique but it is customary to center the volume around the point 
$\Delta p^\alpha=0:$ 
\begin{equation}
P = \int_{\rho^2 g_{\alpha\beta}\, \Delta p^\alpha \Delta \, p^\beta \le r^2(P,n)} 
\frac{{\rho^n\, \sqrt{g}\, \rm d}^n\Delta p}{(2\pi)^{n/2}} 
\exp \left [-\frac{\rho^2}{2}\, g_{\alpha\beta}\, 
\Delta p^\alpha \Delta \, p^\beta \right ].
\label{eq:Vp}
\end{equation}
Given $P$ and the parameter space dimension $n,$ one can iteratively solve 
the above equation for $r(P,n).$ The volume $V_P$ is the surface defined by the equation
\begin{equation}
g_{\alpha\beta} \Delta p^\alpha \Delta p^\beta = 
\left( \frac{r}{\rho} \right)^2.
\label{eq:scaledVolume}
\end{equation}
This is the same as the ellipsoid in Eq.~(\ref{eq:ellipsoid}) except 
that its size is defined by $r/\rho.$ Let us note the generalization of
a result discussed previously, namely that the size of the ellipsoid is not
small enough for all combinations of $P$ and $\rho$ and, therefore, it is
not always possible to interpret the distance from the centre of the ellipsoid 
to its surface in terms of the overlap or match of the signals at the two
locations except when the distance is close to zero. This is because the
expression for the match in terms of the metric is based on the quadratic
approximation which breaks down when the matches are small. However, the 
region defined by Eq.~(\ref{eq:scaledVolume})
always corresponds to the probability $P$ and there is no approximation
here (except that the detector noise is Gaussian). 

When the SNR $\rho$ is large and $1-P$ is not close to zero, the triggers 
are found from the signal with matches greater than or equal to
$1-\tfrac{r^2(P,n)}{2\rho^2}.$ Table \ref{table:one} lists the value of 
$r$ for several values of $P$ in one-, two- and three-dimensions and 
the minimum match $\epsilon_{\rm MM}$ for SNRs 5, 10 and 20.
\begin{table}
\caption{The value of the (squared) distance $d{\ell}^2=r^2/\rho^2$ for several 
values of $P$ and the corresponding smallest match that can be expected between templates 
and the signal at different values of the SNR.}
\label{table:one}
\begin{tabular}{c|cccccccc}
\hline
& \multicolumn{2}{c}{$P=0.683$}& \vline
& \multicolumn{2}{c}{$P=0.954$} & \vline
& \multicolumn{2}{c}{$P=0.997$} \\
\hline
$\rho$ & $d\ell^2$ & $\epsilon_{\rm MM}$ & \vline & $d\ell^2$ & $\epsilon_{\rm MM}$ & \vline &
$d\ell^2$ & $\epsilon_{\rm MM}$ \\
\hline
\multicolumn{8}{c}{$n=1$} \\
\hline
5      & 0.04      & 0.9899  &\vline & 0.16      & 0.9592  &\vline & 0.36      & 0.9055  \\
10     & 0.01      & 0.9975  &\vline & 0.04      & 0.9899  &\vline & 0.09      & 0.9772  \\
20     & 0.0025    & 0.9994  &\vline & 0.01      & 0.9975  &\vline & 0.0225    & 0.9944  \\
\hline
\multicolumn{8}{c}{$n=2$} \\
\hline
5      & 0.092      & 0.9767  &\vline & 0.2470      & 0.9362  &\vline & 0.4800      & 0.8718 \\
10     & 0.023      & 0.9942  &\vline & 0.0618      & 0.9844  &\vline & 0.1200      & 0.9695 \\
20     & 0.00575    & 0.9986  &\vline & 0.0154      & 0.9961  &\vline & 0.0300      & 0.9925 \\
\hline
\multicolumn{8}{c}{$n=3$} \\
\hline
5      & 0.1412     & 0.9641  &\vline & 0.32      & 0.9165  &\vline & 0.568      & 0.8462 \\
10     & 0.0353     & 0.9911  &\vline & 0.08      & 0.9798  &\vline & 0.142      & 0.9638 \\
20     & 0.00883    & 0.9978  &\vline & 0.02      & 0.9950  &\vline & 0.0355     & 0.9911 \\
\hline  
\end{tabular}
\end{table}
Table \ref{table:one} should be interpreted in the light of the fact that
triggers come from an analysis pipeline in which the templates are laid
out with a certain minimal match and one cannot, therefore,
expect the triggers from different detectors to be matched better than the
minimal match. 

From the Table, we see that when the SNR is large (say 
greater than about 10) the dependence of the match $\epsilon_{\rm MM}$ on 
$n$ is very weak; in other words, irrespective of the number of dimensions 
we expect the match between the trigger and the true signal (and for our 
purposes the match between triggers from different instruments) to be 
pretty close to 1, and mostly larger than a minimal match of about $0.95$ 
that is typically used in a search. Even when the SNR is in the region of 
$5,$ for low $P$ again there is a weak dependence 
of $\epsilon_{\rm MM}$ on the number of parameters. For large $P$ and low
SNR, however, the dependence of $\epsilon_{\rm MM}$ on the number of 
dimensions becomes important.  At an SNR of $5$ and $P\simeq 0.997,$ 
$\epsilon_{\rm MM}=0.91, 0.87, 0.85$ for $n=1, 2, 3$ dimensions, respectively.

In general, for a given probability $P$ the size of the ellipsoid at an SNR $\rho$
is smaller by a factor $\rho$ compared to that at $\rho=1.$  Thus, 
the volume in the parameter space in which the measured parameters 
will lie at a given probability $P$ will scale with the SNR as 
$\rho^{-n}$. Therefore, if the goal of an experiment is to have false 
dismissal probability that is no greater than $1-P$ then the 
ellipsoidal windows given by Eq.~(\ref{eq:scaledVolume}) could be employed when testing triggers 
from different detectors for coincidences. We now have our first result which states that:
\begin{quote}
\em {When performing coincidence analysis of triggers one should test to see if
the  associated ellipsoids overlap with each other. 
These ellipsoids describe the smallest possible volume within which the false 
dismissal probability is no more than a pre-specified value.}
\end{quote}
Notice also that, if one assumes that false alarms are due to accidental coincidences between
triggers which are otherwise uncorrelated\footnote{This is not the case for co-located detectors such
as the two LIGO Hanford interferometers.}, the false alarm rate would then also go down by $\rho^{-n}$. Thus, given
the false dismissal probability $1-P$ the size of the ellipsoid further depends on the
SNR of the events that are being subject to coincidence analysis, the size shrinking
sharply as a function of the event's SNR. Thus we have the second of our results:
\begin{quote}
\em {The size of the ellipsoids should be chosen in inverse proportion to the signal-to-noise
ratio.}
\end{quote}
However, this latter feature has not yet been implemented in current
gravitational wave searches and will be a priority for implementation
in future versions of the search pipeline \cite{CBC}.  The final, and practically 
speaking probably the most important, result is the following:
\begin{quote}
\em {Our coincidence algorithm reduces the number of tunable parameters from 
$n$ (where $n$ is the number of parameters) to $1,$ irrespective of the 
dimensionality of the signal parameter space.}
\end{quote}
This parameter $\mu$ introduced in Eq.~\ref{eq:G} essentially scales the 
volume of the ambiguity ellipsoid -- the shape and orientation of which is 
entirely determined from the metric components. The appropriate value of 
this parameter can be determined by extensive Monte-Carlo tests where one 
injects GW signals in interferometer noise and by optimising the detection 
efficiency vis-a-vis false alarm rate, an acceptable value of $\mu$ is 
arrived at. Having just one parameter greatly simplifies this tuning 
procedure. Note that as argued before $\mu$ is SNR-dependent: 
loud signals with high SNRs are expected to be more consistent in 
their parameters in different detectors. Thus the ellipsoids associated 
with these high SNR 
triggers are expected to overlap (and hence pass coincidence) even if 
they each have a smaller volume. On the other hand, for weaker signals 
we need to associate larger ellipsoids in order for them to overlap. 

\section{Overlap of ellipsoids}
\label{subsec:ellipsoidOverlap}

A key tool in determining coincidences of triggers from two or more 
detectors is a mathematical algorithm to determine if the ellipsoids
associated with triggers either touch or overlap with each other.  
This algorithm forms the workhorse for identifying coincidence of 
triggers from two or more detectors. 

As stated in Section \ref{sec2:ScalProd}, triggers resulting from the 
analysis pipeline are projections of the data by normed signal vectors 
onto an $n$-dimensional space ${\cal S}_n,$ where $n$ is the number of 
independent parameters characterizing the source. In the foregoing Section
we introduced ellipsoidal regions in the $n$-dimensional parameter space 
with their centers at the location of the maximum likelihood point. When we analyze the data,
however, we will not know before hand if a signal is present in the data and
even when there is one we would not know where its location in the parameter 
space is. We will have, nevertheless, the knowledge of the location of 
the triggers in the parameter space. Let us denote the coordinates of a
trigger from a detector $A$ as $q^\alpha_A,$ where $\alpha$ is the index 
on the parameter space.  The coincidence analysis
proceeds in the following manner. Define an ellipsoidal region 
${\cal E}({\mathbf p_A},\, {\overline g})$ around each trigger $q^\alpha_A$ by
\begin{equation}
 {\cal E}({\mathbf p_A},\, {\overline g}) = \left \{\mathbf p_A \in {\cal{S}}_n \ \vert \  
          ( \mathbf p_A - \mathbf q_A)^{T} \ \overline g \ (\mathbf p_A - \mathbf q_A) 
\leqslant  1 \right \},
\label{eq:InteriorEllipsoid}
\end{equation}
where $\mathbf q_A \in {\cal{S}}_n$ is the position vector of the center 
of the ellipsoid (i.e., the location of the trigger from detector $A$) and 
$\overline{g}$ is the rescaled metric which we shall refer to as the 
{\it shape matrix}.  It is related to the metric by
\begin{equation}
\overline{g}_{\alpha\beta} = \mu^2 g_{\alpha\beta} 
\label{eq:G}
\end{equation}
where $\mu^2$ is a numerical scaling factor used to expand the linear 
distances of the ellipsoid while holding the position of the center and 
the spatial orientation constant. Eq.~(\ref{eq:scaledVolume}) allows us 
to interpret the parameter $\mu$ in terms of the probability $P$ with which 
the trigger can be expected to be found within the ellipsoid 
${\cal E}({\mathbf p_A},\,{\overline g}):$
\begin{equation}
\mu^2 = \frac {\rho^2} {r^2(P)}.
\end{equation}
Further, the probabilities $P$ associated with a given $\mu$
can be found using Eq.~(\ref{eq:Vp}) when the background noise is
Gaussian. However, most detector noise is non-Gaussian and non-stationary
and in those cases $\mu$ serves as a parameter that must be tuned to achieve
a certain detection efficiency or, alternatively, a certain false alarm rate.

Thus, the shape matrix is 
the scaled metric and encodes the local correlations between 
the parameters in the neighborhood of the trigger center. It is trivial 
to check that when $\mu=1$, Eq.\ (\ref{eq:InteriorEllipsoid}) defines the 
interior of the ambiguity ellipsoid previously defined in Eq.\ (\ref{eq:ambiguity}).

\begin{figure}[htbp!]
\centering
\includegraphics[scale=0.75,angle=0]{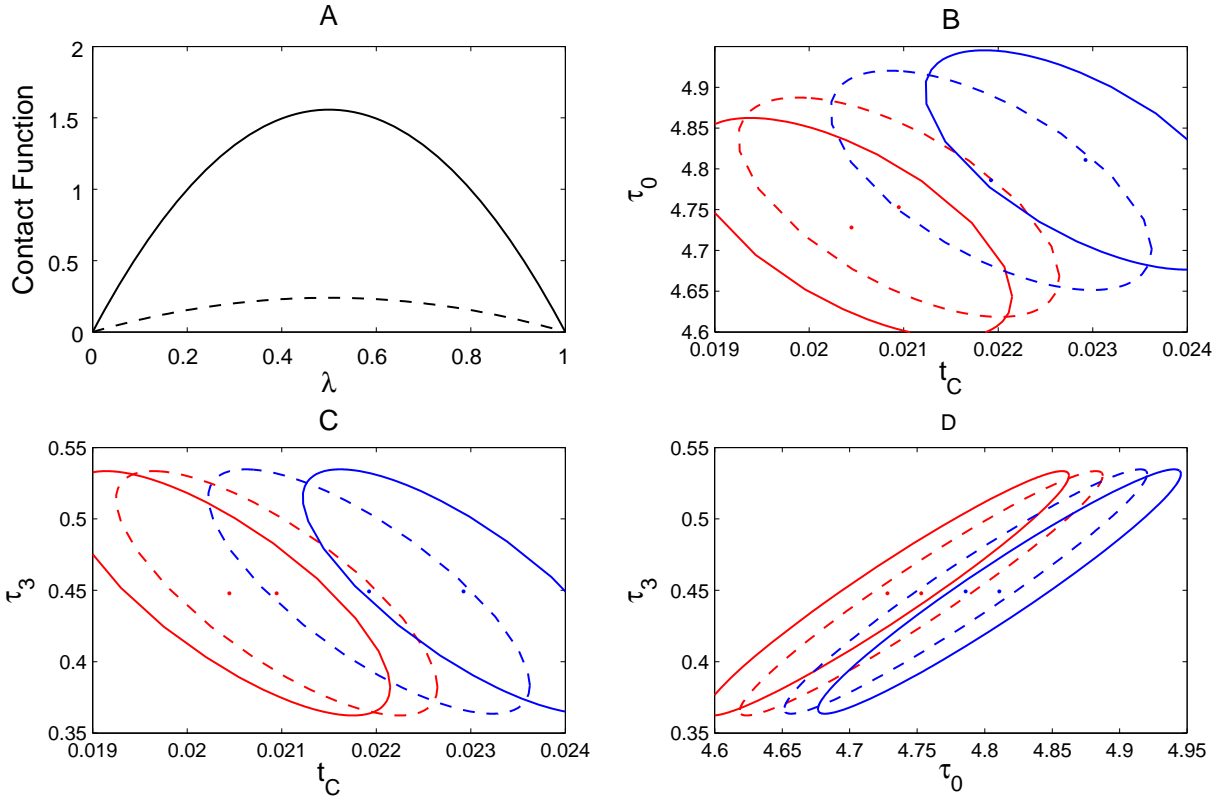}
\caption{Panel A plots the contact function Eq.~(\ref{eq:Contact}) for 
two pairs of three-dimensional ellipsoids taken from a search for
binaries consisting of non-spinning compact objects characterized 
by parameters $(t_c, \tau_0, \tau_3)$ [see Sec.\ \ref{sec3},
in particular Eq.~(\ref{eq:chirptimes})]. Panels B, C and D 
are the projections of the ellipsoids in $(t_c, \tau_0)$, $(t_c, \tau_3)$ 
and $(\tau_0, \tau_3)$ orthogonal planes, respectively. Solid lines refer to 
the case of non-overlapping ellipsoids and dashed lines are for over-lapping 
(i.e., coincident) triggers. Note that in the latter case the maximum of the 
contact function is $\leq 1$, which is the test that is carried out 
to determine if a pair of triggers are in coincidence.}
\label{fig:contactFunction}
\end{figure} 

Once an ellipsoidal model for the trigger is established, following 
\cite{Perram} one can construct a contact function ${\cal F}_{AB}(\lambda)$ 
of two ellipsoids ${\cal E}(\mathbf q_A,\, \overline g_A)$ and 
${\cal E}(\mathbf q_B,\, \overline g_B)$ (defined around triggers from detectors
$A$ and $B$) as 
\begin{equation}
{\cal F}_{AB}(\lambda)  = 
\lambda (1-\lambda) \ \mathbf{r}^{\ T}_{AB} 
\left [ \lambda {\overline{g}}_B^{-1} +  (1-\lambda) 
{\overline{g}}_A^{-1} \right]^{-1} \mathbf{r}_{AB},
\label{eq:Contact}
\end{equation}
where $\mathbf{r}_{AB} = \mathbf q_B - \mathbf q_A$ and $\lambda \in [0,1]$ is 
a scalar parameter.  The maximum of the contact function over $\lambda$
in the interval $[0,1]$ can be shown \cite{Perram} to be unique. 
It can also be shown 
that for two overlapping ellipsoids, the maximum of the contact function 
is less than $1$, i.e,
\begin{equation}
 F = \max_{0 \leqslant \lambda \leqslant 1} \left [  {\cal F}_{AB}(\lambda) \right ] < 1.
\label{eq:maxOverContactFunc}
\end{equation}
When $F=1$, the two ellipsoids 'touch' each other externally.

In the `coincidence' data-analysis paradigm, given triggers from $N$ 
detectors ($N\geq 2$), one draws up a list of `coincident triggers' 
for further analysis to test their significance. The simplest 
coincident triggers consist of those which have `consistent' parameters 
in two detectors (two-way coincidence). Testing for two-way coincidences 
for triggers from co-located detectors (e.g., the two LIGO 
detectors at Hanford) can be accomplished by a single test of 
Eq.~(\ref{eq:maxOverContactFunc}) on a pair of triggers. 

When the detectors are non-colocated, one needs to allow for a 
non-zero `time-of-flight' delay between the trigger arrival times. 
One assumes that the GW signals travel 
at the speed of electro-magnetic radiation in vacuum $c$ and the maximum allowed time 
delay is then set to $\pm \Delta / c,$ where $\Delta$ is the distance between 
the two detectors. As far as the geometrical picture of the coincidence test 
is concerned, for the non-colocated case one needs to test for the overlap 
of a `cylindrical' volume (of length $2 \Delta / c$ along the time-dimension) 
and an ellipsoid\footnote{Note that, in the case of an externally triggered search,
where the position of the source is known, we can use a fixed time delay for non-colocated
detectors.}.
In practice, however, the test can be carried out iteratively 
by adding discrete time delays to the trigger (spanning the allowed 
time-delay) from one detector and testing for the overlap condition against the 
trigger from the other detector, keeping the latter fixed in time. The discrete 
time step can be set to the inverse of the sampling frequency of the 
time series. The fact that the overlap-test is computationally cheap allows for such 
a brute-force implementation strategy to be viable. 

These 2-IFO coincident triggers can now be used as building blocks to 
construct more complex coincidence triggers that have consistent parameters 
over three or more interferometers (3-IFO, ..., $n$-IFO coincidence triggers). For example, the set of triggers $(T_i, T_j, T_k)$ can be classified as 3-IFO coincident if $(T_i, T_j)$, $(T_j, T_k)$ and $(T_i, T_k)$ 2-IFO coincident pairs exist. Here again, the subscripts $i \neq j \neq k$ are labels on interferometers. This idea can be generalised to determine the list of $n$-IFO coincident triggers given the list of $(n-1)$-IFO coincidences. It is useful to note that Eq. \ref{eq:maxOverContactFunc} is the only test we need in order to build the entire hierarchy of coincidence triggers.


We conclude this Section by drawing attention to two practical 
issues in implementing this geometrical coincidence test. The
first has to do with the algorithm one uses to draw up two-way 
coincidences.  Given the set of triggers from two detectors, 
one can (a) work with time-ordered triggers and (b) find the maximum 
length of the bounding box of the ellipsoid along the time dimension over 
all the triggers such that for any trigger from one detector, 
the test for overlap is carried out only if a trigger from the other 
detector occurs at a time that is within twice this interval. 
This approach greatly reduces the overall number of overlap tests 
required to find two-way coincidences. The expression for the length of 
the sides of the bounding box can be algebraically determined given 
the shape matrix of the triggers and is explicitly given for 
2- and 3-dimensions in the next Section.

The second point is on the numerical implementation of the test of the 
overlap of ellipsoids where we maximize the contact function over a single 
parameter $\lambda$. Evaluation of the contact function involves matrix 
inversion which can be quite expensive computationally. Under these 
circumstances, prior  knowledge of the inverse of trigger shape matrices 
can prove to be more efficient than on-the-fly computation. Brent's 
minimization method \cite{WikiBrent,BookBrent} is particularly suitable 
for fast convergence to the maxima given the well behaved nature of the 
contact function and is available as part of the GNU Scientific Library \cite{GSL}.

\section{Expected reduction in false alarm rate}
\label{sec:reduction in FAR}
Next, let us consider the reduction in the false alarm rate as a result of using
ellipsoidal windows as opposed to rectangular windows\footnote{This discussion again 
assumes that false alarms are due to accidental coincidences between otherwise uncorrelated
triggers.}. In order to achieve 
false dismissal probability less than or equal to $1-P,$ a rectangular window has
to be at least as large as the box that encloses the ellipsoid. Now the volume
of an $n$-dimensional ellipsoid ($n\ge 2$) whose semi-axes are $a_k,$ $k=1,
\ldots, n,$ is given by a recursive formula:
\begin{equation}
V_{n} = \frac{ 2 \pi V_{n-2}}{n} \, \prod_{k=1}^n a_k,\ \ {\rm where\ } V_0=1,\ \ V_1=2.
\end{equation}
On the other hand, the smallest volume an $n$-dimensional box that encloses the 
ellipsoid would be
\begin{equation}
U_{n} = \prod_{k=1}^n (2 a_k) = 2^n \prod_{k=1}^n a_k,
\end{equation}
where a factor of 2 arises since $a_k$ are semi-major axes and the side-lengths of
the enclosing box will be twice that value. Thus, the rectangular box's volume 
is larger than that of the ellipsoid by the factor
\begin{equation}
r \equiv \frac{U_n}{V_n} = \frac{n\,2^{n-1}}{\pi V_{n-2}}.
\end{equation}
Thus, in 2-, 3- and 4-dimensions the saving is $4/\pi,$  $6/\pi$ and $32/\pi^2,$
respectively. However, the real factor could be far greater as the error
ellipsoids are generally not oriented along the coordinate axes. 

When the ellipsoid is not aligned with the coordinate axes, which will be the case
when there are correlations between the different parameters, the
side-lengths of the bounding box are given by maximizing each coordinate
axis over the entire ellipsoidal surface as follows. Starting from
Eq.~(\ref{eq:scaledVolume}) one can express the first of the coordinates
$p \equiv p^1$ in terms of the other coordinates:
\begin{equation}
g_{11} p^2 + 2 g_{1i} p\, p^i + g_{ij} p^i p^j - \left (\frac{r}{\rho} \right )^2 = 0,
\ \ \ i, j= 2, \ldots n,
\label{eq:quadratic}
\end{equation}
which can be solved to obtain
\begin{equation}
p_{\pm} = \frac{1}{g_{11}} \left [
- g_{1i}\,p^i \pm \sqrt { \left ( g_{1i}\, g_{1j}  - 
g_{11}\, g_{ij} \right) p^i p^j + (g_{11}\, r^2/\rho^2)}
\right ].
\end{equation}
For our purposes we only need the `plus' solution.
One can then set-up $n-1$ equations in as many variables by demanding that
$\partial p_+/ \partial p^k = 0,$ which gives
\begin{equation}
\left[\frac{(g_{1i}g_{1k}-g_{11}g_{ik})(g_{1j}g_{1k}-g_{11}g_{jk})}{g_{1k}^2}
- (g_{1i}g_{1j}-g_{11}g_{ij})\right]p^i p^j = g_{11}\frac{r^2}{\rho^2}.
\end{equation}
These are again quadratic equations that must be solved (simultaneously)
for the coordinates $p^j,$ $j=2\,\ldots,\,n.$ The resulting (positive)
roots, denoted $p_1^j$ can be substituted in Eq.~(\ref{eq:quadratic})
to obtain the half-side-length of the ellipse. We shall next
give explicit expressions for the side-lengths of the enclosing box in
two and three dimensions.  In higher dimensions the expressions
are rather cumbersome but the general procedure outlined above 
can be used to compute the volume of the bounding box in all cases.

The side-lengths of the bounding box are given in two dimensions by 
\begin{eqnarray}
\label{eq:2dBoundingBox}
x = 2 \sqrt{\frac{\G_{22}}{|\G|}}, \ \ y = 2 \sqrt{\frac{\G_{11}}{|\G|}},
\end{eqnarray}
and in three dimensions by:
\begin{eqnarray}
\label{eq:3dBoundingBox}
x = 2 \sqrt{ \frac{\left( \G_{23}^2 - \G_{22}\G_{33}\right) \G_{22}}
{\left( \G_{12}\G_{23} - \G_{22}\G_{13}\right)^2 - \left(\G_{23}^2 -
\G_{22}\G_{33}\right)\left(\G_{12}^2 - \G_{11}\G_{22}\right) }},\nonumber \\
y = 2 \sqrt{ \frac{\left( \G_{13}^2 - \G_{11}\G_{33}\right) \G_{11}}
{\left(\G_{12}\G_{13} - \G_{11}\G_{23}\right)^2 - \left(\G_{13}^2 -
\G_{11}\G_{33}\right) \left(\G_{12}^2 - \G_{11}\G_{22}\right)}},\nonumber \\
z = 2 \sqrt{ \frac{\left( \G_{12}^2 - \G_{11}\G_{22}\right) \G_{11}}
{\left(\G_{12}\G_{13} - \G_{11}\G_{23}\right)^2 - \left(\G_{12}^2 -
\G_{11}\G_{22}\right) \left(\G_{13}^2 - \G_{11}\G_{33}\right)}}.
\end{eqnarray}

\section{Application to coalescing binaries}
\label{sec3}

Inspiralling compact binaries are one of the most promising candidates 
for detection by the laser interferometric 
detectors. It will, therefore, be interesting to investigate the gains 
of using the new coincidence method in such searches. For the purpose 
of our discussion, it will suffice to use a simple model of the signal. 
We shall use the Fourier representation of the waveform from a binary
consisting of non-spinning compact objects on a quasi-circular orbit
in which post-Newtonian (PN) corrections to the amplitude are 
neglected, but corrections to the phase are included to the desired 
order. This waveform is calculated using the stationary phase 
approximation, and is of the form:
\begin{equation}
\label{eq:SPA}
\tilde{h}(f) = \frac{AM^{5/6}}{D\pi^{2/3}}\sqrt{\frac{5 \eta}{24}} f^{-7/6} \exp
\left[i \Psi(f; t_C, \phi_C, k) + i\frac{\pi}{4}\right],
\end{equation}
\begin{equation}
\label{SPAPhase}
\Psi(f) = 2 \pi f t_C + \phi_C  + \sum_k \lambda_k f^{(k-5)/3}
\end{equation}
Where $D$ is the distance to the source, and $A$ is a constant which depends on
the relative orientations of the detector and the binary orbit, and $t_C$ and $\phi_C$
are as defined in Section~\ref{sec2:ScalProd}.
Waveforms of this type at 2PN order \cite{BDI95,BDIWW95} have been used 
in previous searches for binary neutron star inspirals~\cite{Abbott:2005pe}, 
and are currently being used in searches for compact binary inspirals 
with a total mass of $< 35 M_\odot$ \cite{CBC}. Moreover, the metric computed for 
such a waveform has been shown to be approximately valid for a range 
of physical approximants~\cite{Bank06,cokelaer:102004}. At the 2PN order, the 
coefficients $\lambda_k$ are given by the following expressions:
\begin{eqnarray}
\lambda_0 = \frac{3}{128 \eta(\pi M)^{5/3}}, \ \ \lambda_1 = 0, \ \ 
\lambda_2 = \frac{5}{96 \pi \eta M}\left(\frac{743}{336} 
+ \frac{11}{4}\eta \right), \nonumber \\
\lambda_3 = \frac{-3 \pi^{1/3}}{8 \eta M^{2/3}}, \ \ 
\lambda_4 = \frac{15}{64 \eta (\pi M^{1/3}}\left(\frac{3058673}{1016064} 
+ \frac{5429}{1008}\eta + \frac{617}{144}\eta^2\right),
\end{eqnarray}
where $M$ is the total mass of the system, and $\eta$ is the 
symmetric mass ratio, which is defined as $\eta \equiv m_1 m_2 / M^2$.

The metric required for determining coincidence in the case of 
non-spinning binaries is that in the 3-dimensional space of 
$(t_C, \tau_0, \tau_3)$, where $\tau_0$ and $\tau_3$ are the chirp 
times, which are a convenient way of parameterizing the masses of the 
binary system. They are given by
\begin{eqnarray}
\label{eq:chirptimes}
\tau_0 = \frac{5}{256 \pi f_L \eta}(\pi M f_L)^{-5/3}, \ \ 
\tau_3 = \frac{1}{8 f_L \eta}(\pi M f_L)^{-2/3},
\end{eqnarray}
where $f_L$ is the frequency below which no appreciable signal can be detected due
to rising detector noise at low frequencies.

In obtaining the metric, it proves to be more
convenient to use parameters $(t_C, \theta_1, \theta_2)$, where 
$\theta_1 \equiv 2 \pi f_L \tau_0$, and $\theta_2 \equiv 2\pi f_L \tau_3$.
This metric was obtained by Owen in \cite{Owen96}. Here, Eq.~(\ref{eq:overlap}) 
was used, and the phase $\phi_C$ maximized over to give the expression for the 
metric:
\begin{equation}
\label{eq:metricExpression}
g_{\alpha \beta} = \frac{1}{2} \left(\mathcal{J}\left[\psi_\alpha \psi_\beta \right] - 
\mathcal{J}\left[\psi_\alpha \right] \mathcal{J}\left[ \psi_\beta \right] \right),
\end{equation}
where $\psi_\alpha$ is the derivative of the Fourier phase of the inspiral 
waveform with respect to parameter $\theta_\alpha$. $\mathcal{J}$ is the 
moment functional of the noise PSD, which is defined for any function $a(x)$ as:
\begin{equation}
\mathcal{J}(a) \equiv  \frac{1}{I(7)} \int^{x_U}_{x_L} 
\frac{a(x)x^{-7/3}}{S_h(x)} dx.
\end{equation}
$I(q)$ is the $q$th moment of the noise PSD, which is defined by:
\begin{equation}
I(q) \equiv S_h(f_0) \int^{x_U}_{x_L} \frac{x^{-q/3}}{S_h(x)} dx,
\end{equation}
where $x \equiv f/f_0$, $f_0$ being a fiducial frequency used to set the 
range of the numerical values of the functions contained in the 
integrals. The value of $x_L$ is chosen so that the contribution to 
the integral for values below $x_L$ would be negligible. 
$x_U \equiv f_U / f_0$, where $f_U$ is the ending frequency of the 
inspiral waveform in question.  In deriving the explicit expression 
for the metric, the starting point is the Fourier phase of the 
waveform in the form \cite{Bank06}:
\begin{eqnarray}
\Psi(f; t_C, \theta_1, \theta_2)  & = & 2 \pi f t_C + a_{01} \theta_1 x^{-5/3} + 
\left[a_{21} \left(\theta_1/\theta_2 \right ) 
+ a_{22}\left (\theta_1 \theta_2^2\right)^{1/3}\right] x^{-1} 
+ a_{31} \theta_2 x^{-2/3}  \nonumber \\
& + & \left[a_{41}\left (\theta_1/\theta_2^2 \right ) 
+ a_{42}\left(\theta_1/\theta_2\right)^{1/3} +
a_{43}\left(\theta_2^4/\theta_1 \right)^{1/3}\right] x^{-1/3},
\end{eqnarray}
where the coefficients $a_{km}$ are given by
\begin{eqnarray}
a_{01}  =  \frac{3}{5},\ \  a_{21} = \frac{11 \pi}{12},\ \  
a_{22} = \frac{743}{2016}\left(\frac{25}{2 \pi^2}\right)^{1/3}, \ \ 
a_{31} = \frac{-3}{2},\ \  a_{41} = \frac{617}{384}  \pi^2, \nonumber \\
a_{42} = \frac{5429}{5376}\left(\frac{25 \pi}{2}\right)^{1/3}  \ \
a_{43}  =  \frac{15293365}{10838016} \left(\frac{5}{4\pi^4}\right)^{1/3}
\end{eqnarray}
Using the above in Eq.~(\ref{eq:metricExpression}), one can find 
an explicit expression for the metric.  This expression is too unwieldy 
to write here, but it can be obtained by utilising the fact that, since 
the Fourier phase is a polynomial function, $\mathcal{J}$ can be 
expanded in terms of normalised moments $J$, where
\begin{equation}
J(p) \equiv \frac{I(p)}{I(7)}.
\end{equation}

To assess the potential gains of using this coincidence method for inspiral 
analysis, it is useful to consider the difference in volume between the 
ellipsoidal region defined by $\G$, and its bounding box aligned with the 
co-ordinate axes $(t_C, \tau_0, \tau_3)$. This ratio can be calculated 
with the help of Eqs.~(\ref{eq:3dBoundingBox}).
\begin{figure}
\centering
\includegraphics[scale=0.2,angle=0]{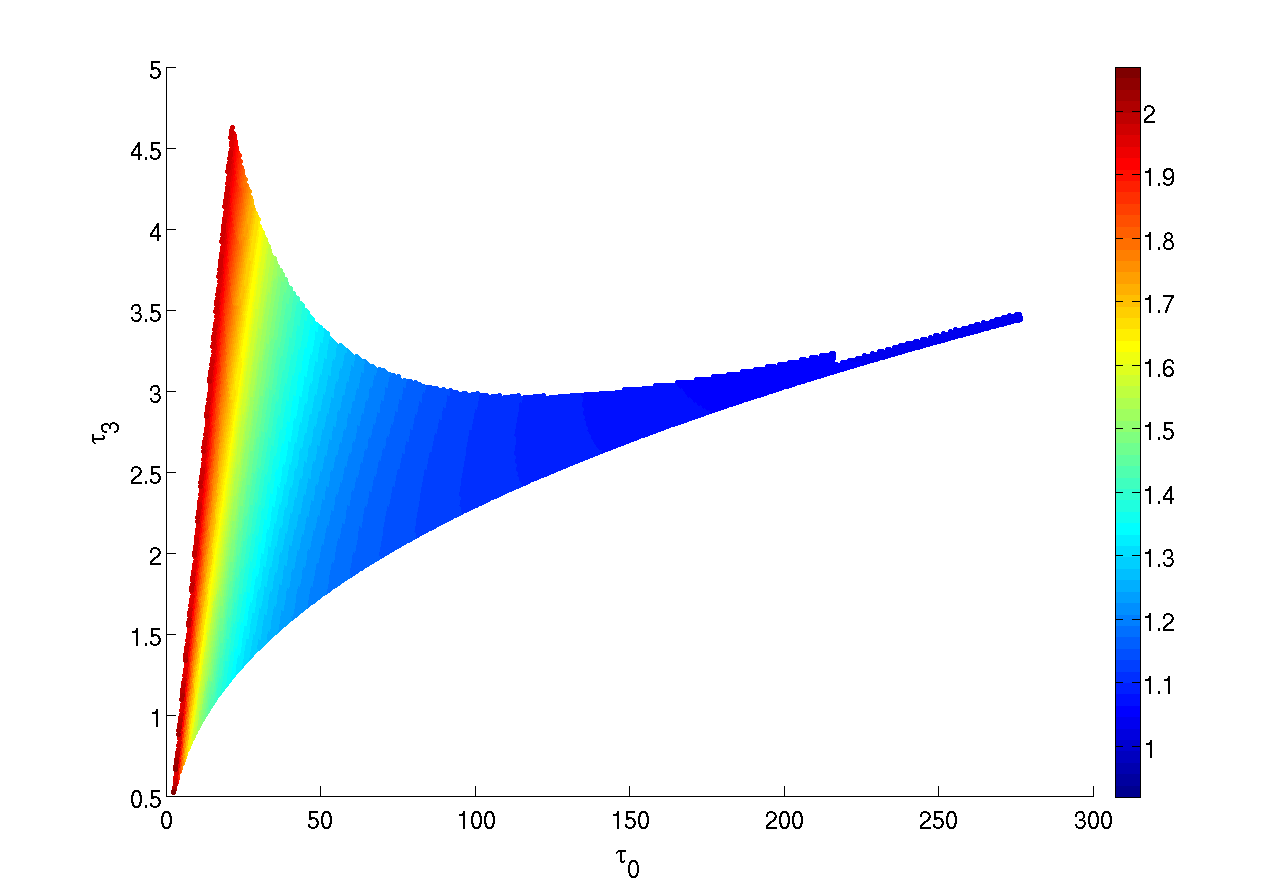}
\includegraphics[scale=0.2,angle=0]{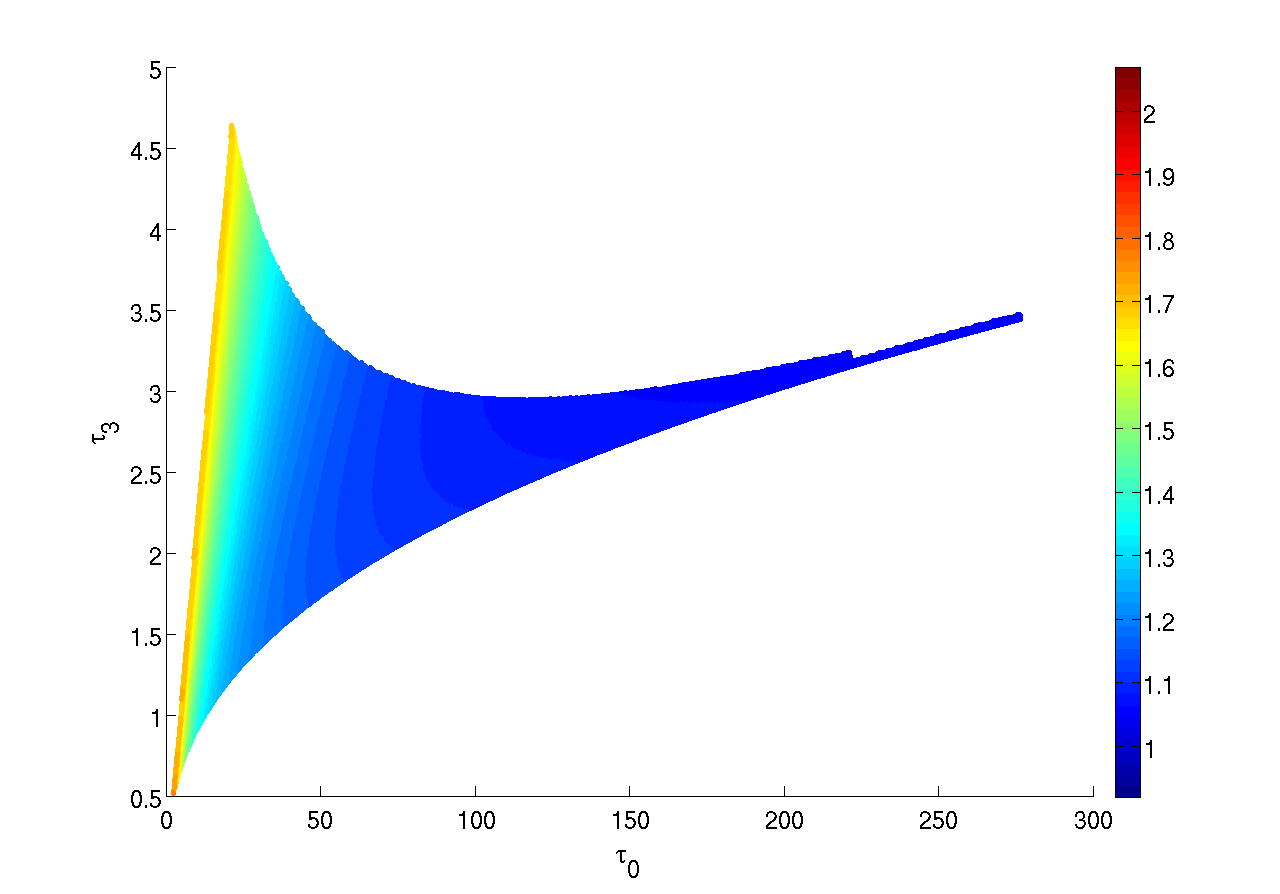}
\includegraphics[scale=0.2,angle=0]{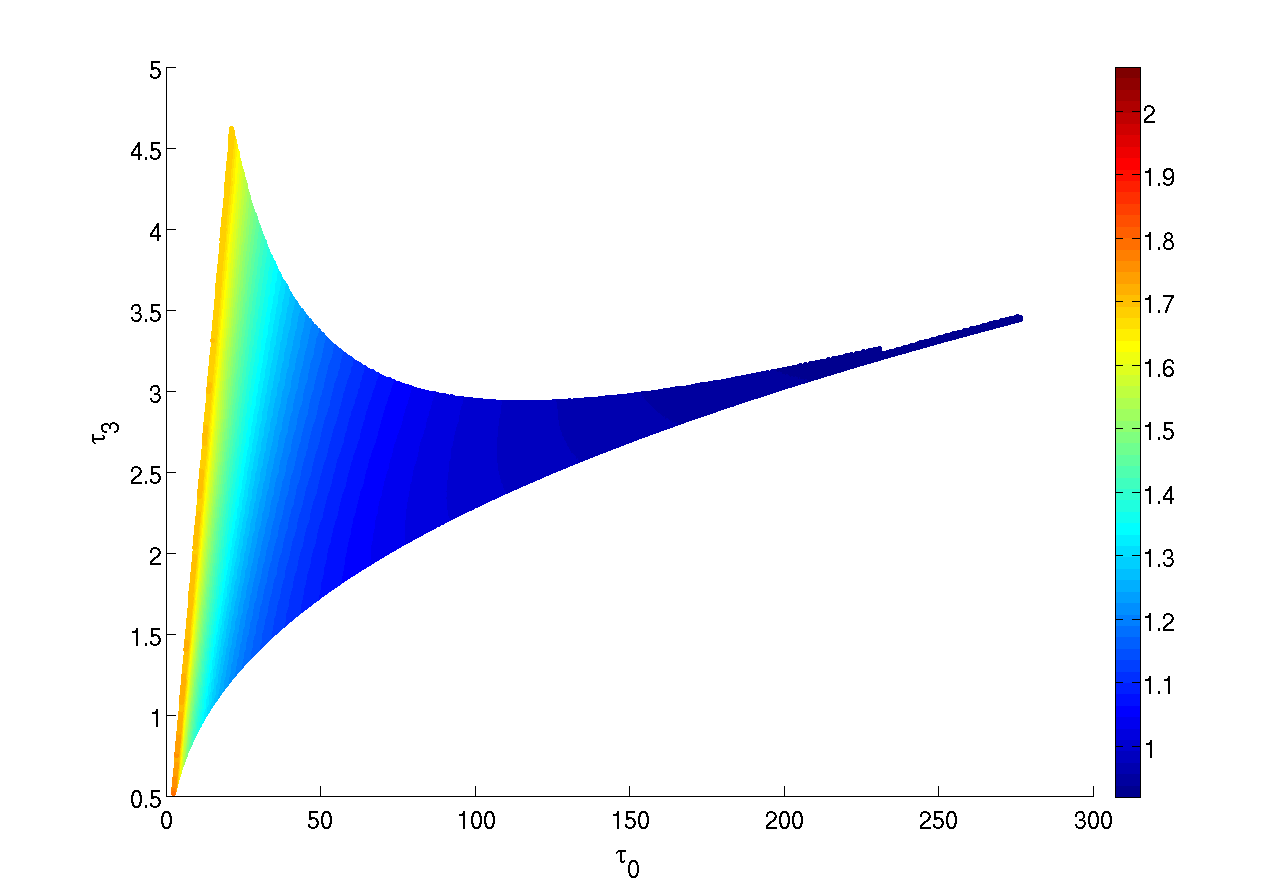}
\includegraphics[scale=0.2,angle=0]{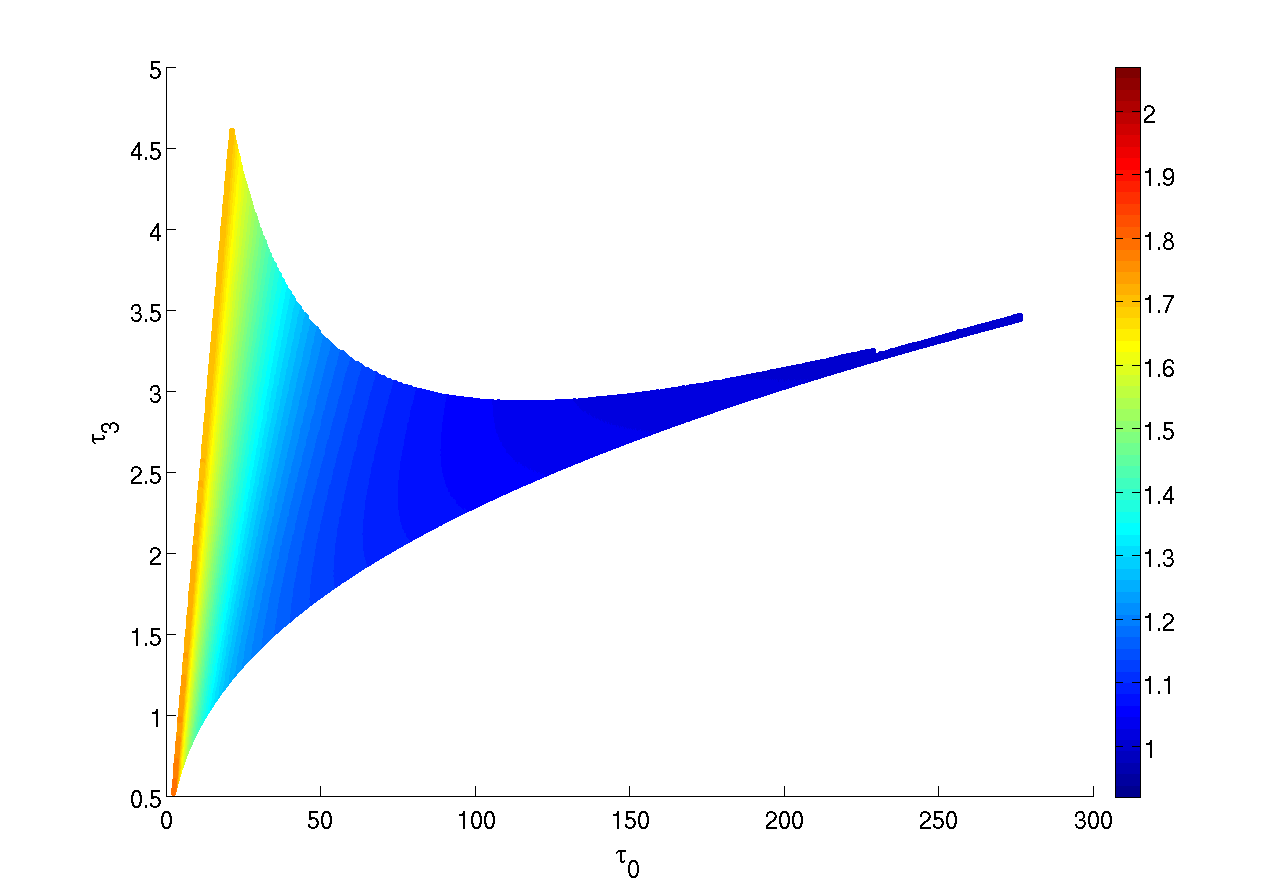}
\caption{The $\log_{10}$ of the ratio of the volume of the bounding box 
to the volume of the ellipsoid as a function of location in 
$(\tau_0, \tau_3)$ space. The plots shown are (clockwise from top-right) 
for the initial LIGO, advanced LIGO, Virgo and Einstein Telescope.  
The low frequency cutoff is chosen to be $20$ Hz.}
\label{fig:ratioBoxEllipsoid}
\end{figure}
Fig.~\ref{fig:ratioBoxEllipsoid} shows how this ratio varies 
across the $(\tau_0, \tau_3)$ space in the case of Initial and Advanced
LIGO, Virgo and Einstein Telescope
(a third generation European detector that is currently being designed).
It can be seen that for most of the parameter space, the volume of the
bounding box is an order of magnitude larger than the volume of the
ellipsoid; however, in certain regions, corresponding to high masses, this
ratio can be as large as two orders of magnitude. 
This suggests that significant reductions of the background can be 
achieved by using ellipsoidal windows.  Runs on example data sets suggest 
that in practice, the reduction in background coincident triggers due 
to using such a coincidence method will be a factor of $\sim 10$.

To assess the improvement in the confidence in any candidate detection,
it is helpful to look at how reducing the background rate by a factor of $k$
will improve the odds $O$ of a detection
\begin{equation}
\label{eq:odds}
O(h|D) = \frac{P(h|D)}{P(0|D)},
\end{equation}
where $P(h|D)$ is the posterior probability of a signal $h$ being present given the
set of triggers $D$ has been obtained, and $P(0|D)$ is the probability of there 
being no signal given $D.$ We take the accidental trigger rate
to be a Poisson process, with a trigger rate prior to reduction $\lambda$.
Assuming that the detection efficiency is not affected by the reduction in the trigger
rate, we see that the odds improves by the following factor:
\begin{equation}
\label{eq:oddsImprovement}
\frac{O(h|D)_{\lambda/k}}{O(h|D)_{\lambda}} = \frac{1-e^{-\lambda T}}{1-e^{-\lambda T/k}},
\end{equation}
where $T$ is the duration of the run. 

The factor by which the odds of a signal being present improves
by reducing the false alarm rate by a factor of $k$ depends on how high the false
alarm rate was to start with. If the initial false alarm rate is low $(\lambda T \ll 1)$,
the improvement in the odds approaches the factor $k$. However, for high false alarm
rates, the improvement becomes less marked, tending to a factor of 1 
as $\lambda T \rightarrow \infty$.

\section{Summary and Conclusions}
\label{sec4}
A new method of coincidence analysis is proposed in which, instead of the 
rectangular windows on parameters conventionally used, ellipsoidal 
windows are employed based on the metric defined on the signal manifold. This allows 
us to use windows of appropriate size depending on the location in the 
parameter space, instead of using a phenomenological `best fit' choice of 
windows across the entire space. The algorithm has a massive practical
advantage in that it requires the tuning of only one parameter irrespective 
of the number of dimensions of the parameters.  This contrasts with the
conventional method that required us to tune nearly as many parameters
as the dimension of the parameter space. In addition, the method allows us to take 
into account covariances between parameters, thus significantly reducing 
the volume enclosed within the windows. In particular, for the case of 
non-spinning compact binary coalescence in Initial LIGO, it is expected 
that the use of such a method will reduce the background rate of coincident 
triggers by roughly an order of magnitude. By also incorporating 
SNR-dependence into the size of the windows, the background of high SNR 
events can be reduced even further.  

The algorithm has been implemented in C code in the LSC Algorithm Library (LAL) \cite{LAL}.
An implementation in $(t_C, \tau_0, \tau_3)$ space, as in Section~\ref{sec3}, using SNR-independent
windows, is being employed in the search for compact binary coalescence in S5 data. This implementation
is referred to as \emph{e-thinca} \cite{CBC}.

\acknowledgments
The authors gratefully acknowledge the members of the Compact Binary Coalescence working group of the LIGO Scientific Collaboration for their contributions, suggestions and discussions regarding this work. The authors are also grateful to Alan Weinstein, Michele Valisneri, Alexander Dietz and Duncan Brown for carefully reading the manuscript and for suggesting improvements to the text.

\phantomsection
\addcontentsline{toc}{chapter}{Bibliography}
\bibliography{ref-list}

\begin{thebibliography}{47}
\expandafter\ifx\csname natexlab\endcsname\relax\def\natexlab#1{#1}\fi
\expandafter\ifx\csname bibnamefont\endcsname\relax
  \def\bibnamefont#1{#1}\fi
\expandafter\ifx\csname bibfnamefont\endcsname\relax
  \def\bibfnamefont#1{#1}\fi
\expandafter\ifx\csname citenamefont\endcsname\relax
  \def\citenamefont#1{#1}\fi
\expandafter\ifx\csname url\endcsname\relax
  \def\url#1{\texttt{#1}}\fi
\expandafter\ifx\csname urlprefix\endcsname\relax\def\urlprefix{URL }\fi
\providecommand{\bibinfo}[2]{#2}
\providecommand{\eprint}[2][]{\url{#2}}

\bibitem[{\citenamefont{Abbott et~al.}(2004{\natexlab{a}})}]{Abbott:2003vs}
\bibinfo{author}{\bibfnamefont{B.}~\bibnamefont{Abbott}} \bibnamefont{et~al.}
  (\bibinfo{collaboration}{LIGO Scientific}), \bibinfo{journal}{Nucl. Instrum.
  Meth.} \textbf{\bibinfo{volume}{A517}}, \bibinfo{pages}{154}
  (\bibinfo{year}{2004}{\natexlab{a}}), \eprint{gr-qc/0308043}.

\bibitem[{\citenamefont{Acernese et~al.}(2006)}]{Acernese:2006bj}
\bibinfo{author}{\bibfnamefont{F.}~\bibnamefont{Acernese}}
  \bibnamefont{et~al.}, \bibinfo{journal}{Class. Quant. Grav.}
  \textbf{\bibinfo{volume}{23}}, \bibinfo{pages}{S635} (\bibinfo{year}{2006}).

\bibitem[{\citenamefont{Luck et~al.}(2006)}]{Luck:2006ug}
\bibinfo{author}{\bibfnamefont{H.}~\bibnamefont{Luck}} \bibnamefont{et~al.},
  \bibinfo{journal}{Class. Quant. Grav.} \textbf{\bibinfo{volume}{23}},
  \bibinfo{pages}{S71} (\bibinfo{year}{2006}).

\bibitem[{\citenamefont{Pai et~al.}(2001)\citenamefont{Pai, Dhurandhar, and
  Bose}}]{Pai:2000zt}
\bibinfo{author}{\bibfnamefont{A.}~\bibnamefont{Pai}},
  \bibinfo{author}{\bibfnamefont{S.}~\bibnamefont{Dhurandhar}},
  \bibnamefont{and} \bibinfo{author}{\bibfnamefont{S.}~\bibnamefont{Bose}},
  \bibinfo{journal}{Phys. Rev.} \textbf{\bibinfo{volume}{D64}},
  \bibinfo{pages}{042004} (\bibinfo{year}{2001}), \eprint{gr-qc/0009078}.

\bibitem[{\citenamefont{Bose et~al.}(1999)\citenamefont{Bose, Dhurandhar, and
  Pai}}]{Bose:1999bp}
\bibinfo{author}{\bibfnamefont{S.}~\bibnamefont{Bose}},
  \bibinfo{author}{\bibfnamefont{S.~V.} \bibnamefont{Dhurandhar}},
  \bibnamefont{and} \bibinfo{author}{\bibfnamefont{A.}~\bibnamefont{Pai}},
  \bibinfo{journal}{Pramana} \textbf{\bibinfo{volume}{53}},
  \bibinfo{pages}{1125} (\bibinfo{year}{1999}), \eprint{gr-qc/9906064}.

\bibitem[{\citenamefont{Finn}(2001)}]{Finn:2000hj}
\bibinfo{author}{\bibfnamefont{L.~S.} \bibnamefont{Finn}},
  \bibinfo{journal}{Phys. Rev.} \textbf{\bibinfo{volume}{D63}},
  \bibinfo{pages}{102001} (\bibinfo{year}{2001}), \eprint{gr-qc/0010033}.

\bibitem[{\citenamefont{Arnaud et~al.}(2003)}]{Arnaud:2003zq}
\bibinfo{author}{\bibfnamefont{N.}~\bibnamefont{Arnaud}} \bibnamefont{et~al.},
  \bibinfo{journal}{Phys. Rev.} \textbf{\bibinfo{volume}{D68}},
  \bibinfo{pages}{102001} (\bibinfo{year}{2003}), \eprint{gr-qc/0307100}.

\bibitem[{\citenamefont{Jaranowski and Krolak}(1994)}]{Jaranowski:1994xd}
\bibinfo{author}{\bibfnamefont{P.}~\bibnamefont{Jaranowski}} \bibnamefont{and}
  \bibinfo{author}{\bibfnamefont{A.}~\bibnamefont{Krolak}},
  \bibinfo{journal}{Phys. Rev.} \textbf{\bibinfo{volume}{D49}},
  \bibinfo{pages}{1723} (\bibinfo{year}{1994}).

\bibitem[{\citenamefont{Jaranowski and Krolak}(1996)}]{Jaranowski:1994}
\bibinfo{author}{\bibfnamefont{P.}~\bibnamefont{Jaranowski}} \bibnamefont{and}
  \bibinfo{author}{\bibfnamefont{A.}~\bibnamefont{Krolak}},
  \bibinfo{journal}{Class. Quantum Grav.} \textbf{\bibinfo{volume}{13}},
  \bibinfo{pages}{1279} (\bibinfo{year}{1996}).

\bibitem[{\citenamefont{Tagoshi et~al.}(2007)}]{Tagoshi:2007ni}
\bibinfo{author}{\bibfnamefont{H.}~\bibnamefont{Tagoshi}} \bibnamefont{et~al.},
  \bibinfo{journal}{Phys. Rev.} \textbf{\bibinfo{volume}{D75}},
  \bibinfo{pages}{087306} (\bibinfo{year}{2007}), \eprint{gr-qc/0702019}.

\bibitem[{\citenamefont{Abbott et~al.}(2004{\natexlab{b}})}]{Abbott:2003pj}
\bibinfo{author}{\bibfnamefont{B.}~\bibnamefont{Abbott}} \bibnamefont{et~al.}
  (\bibinfo{collaboration}{LIGO Scientific}), \bibinfo{journal}{Phys. Rev.}
  \textbf{\bibinfo{volume}{D69}}, \bibinfo{pages}{122001}
  (\bibinfo{year}{2004}{\natexlab{b}}), \eprint{gr-qc/0308069}.

\bibitem[{\citenamefont{Abbott et~al.}(2005{\natexlab{a}})}]{Abbott:2005pe}
\bibinfo{author}{\bibfnamefont{B.}~\bibnamefont{Abbott}} \bibnamefont{et~al.}
  (\bibinfo{collaboration}{LIGO Scientific}), \bibinfo{journal}{Phys. Rev.}
  \textbf{\bibinfo{volume}{D72}}, \bibinfo{pages}{082001}
  (\bibinfo{year}{2005}{\natexlab{a}}), \eprint{gr-qc/0505041}.

\bibitem[{\citenamefont{Abbott et~al.}(2005{\natexlab{b}})}]{Abbott:2005pf}
\bibinfo{author}{\bibfnamefont{B.}~\bibnamefont{Abbott}} \bibnamefont{et~al.}
  (\bibinfo{collaboration}{LIGO Scientific}), \bibinfo{journal}{Phys. Rev.}
  \textbf{\bibinfo{volume}{D72}}, \bibinfo{pages}{082002}
  (\bibinfo{year}{2005}{\natexlab{b}}), \eprint{gr-qc/0505042}.

\bibitem[{\citenamefont{Abbott et~al.}(2006{\natexlab{a}})}]{Abbott:2005kq}
\bibinfo{author}{\bibfnamefont{B.}~\bibnamefont{Abbott}} \bibnamefont{et~al.}
  (\bibinfo{collaboration}{LIGO Scientific}), \bibinfo{journal}{Phys. Rev.}
  \textbf{\bibinfo{volume}{D73}}, \bibinfo{pages}{062001}
  (\bibinfo{year}{2006}{\natexlab{a}}), \eprint{gr-qc/0509129}.

\bibitem[{\citenamefont{Arnaud et~al.}(2002)}]{Arnaud:2001my}
\bibinfo{author}{\bibfnamefont{N.}~\bibnamefont{Arnaud}} \bibnamefont{et~al.},
  \bibinfo{journal}{Phys. Rev.} \textbf{\bibinfo{volume}{D65}},
  \bibinfo{pages}{042004} (\bibinfo{year}{2002}), \eprint{gr-qc/0107081}.

\bibitem[{\citenamefont{Mukhopadhyay et~al.}(2006)\citenamefont{Mukhopadhyay,
  Sago, Tagoshi, Dhurandhar, Takahashi, and Kanda}}]{CoherentvsCoincident}
\bibinfo{author}{\bibfnamefont{H.}~\bibnamefont{Mukhopadhyay}},
  \bibinfo{author}{\bibfnamefont{N.}~\bibnamefont{Sago}},
  \bibinfo{author}{\bibfnamefont{H.}~\bibnamefont{Tagoshi}},
  \bibinfo{author}{\bibfnamefont{S.}~\bibnamefont{Dhurandhar}},
  \bibinfo{author}{\bibfnamefont{H.}~\bibnamefont{Takahashi}},
  \bibnamefont{and} \bibinfo{author}{\bibfnamefont{N.}~\bibnamefont{Kanda}},
  \bibinfo{journal}{Phys. Rev.} \textbf{\bibinfo{volume}{D74}},
  \bibinfo{pages}{083005} (\bibinfo{year}{2006}), \eprint{gr-qc/0608103}.

\bibitem[{\citenamefont{Abbott et~al.}(2005{\natexlab{c}})}]{Abbott:2005fb}
\bibinfo{author}{\bibfnamefont{B.}~\bibnamefont{Abbott}} \bibnamefont{et~al.}
  (\bibinfo{collaboration}{LIGO Scientific}), \bibinfo{journal}{Phys. Rev.}
  \textbf{\bibinfo{volume}{D72}}, \bibinfo{pages}{062001}
  (\bibinfo{year}{2005}{\natexlab{c}}), \eprint{gr-qc/0505029}.

\bibitem[{\citenamefont{Abbott et~al.}(2006{\natexlab{b}})}]{Abbott:2005at}
\bibinfo{author}{\bibfnamefont{B.}~\bibnamefont{Abbott}} \bibnamefont{et~al.}
  (\bibinfo{collaboration}{LIGO Scientific}), \bibinfo{journal}{Class. Quant.
  Grav.} \textbf{\bibinfo{volume}{23}}, \bibinfo{pages}{S29}
  (\bibinfo{year}{2006}{\natexlab{b}}), \eprint{gr-qc/0511146}.

\bibitem[{\citenamefont{Cannon}(2007)}]{Cannon:2008}
\bibinfo{author}{\bibfnamefont{K.}~\bibnamefont{Cannon}}
  (\bibinfo{year}{2007}), \bibinfo{note}{g070881-00-Z}.

\bibitem[{\citenamefont{Cutler and Flanagan}(1994)}]{CF94}
\bibinfo{author}{\bibfnamefont{C.}~\bibnamefont{Cutler}} \bibnamefont{and}
  \bibinfo{author}{\bibfnamefont{E.}~\bibnamefont{Flanagan}},
  \bibinfo{journal}{Phys. Rev. D} \textbf{\bibinfo{volume}{49}},
  \bibinfo{pages}{2658} (\bibinfo{year}{1994}).

\bibitem[{\citenamefont{Finn}(1992)}]{Finn92}
\bibinfo{author}{\bibfnamefont{L.}~\bibnamefont{Finn}}, \bibinfo{journal}{Phys.
  Rev. D} \textbf{\bibinfo{volume}{46}}, \bibinfo{pages}{5236}
  (\bibinfo{year}{1992}).

\bibitem[{\citenamefont{Finn and Chernoff}(1993)}]{FinnCh93}
\bibinfo{author}{\bibfnamefont{L.}~\bibnamefont{Finn}} \bibnamefont{and}
  \bibinfo{author}{\bibfnamefont{D.}~\bibnamefont{Chernoff}},
  \bibinfo{journal}{Phys. Rev. D} \textbf{\bibinfo{volume}{47}},
  \bibinfo{pages}{2198} (\bibinfo{year}{1993}).

\bibitem[{\citenamefont{Chernoff and Finn}(1993)}]{Chernoff:1993th}
\bibinfo{author}{\bibfnamefont{D.~F.} \bibnamefont{Chernoff}} \bibnamefont{and}
  \bibinfo{author}{\bibfnamefont{L.~S.} \bibnamefont{Finn}},
  \bibinfo{journal}{Astrophys. J.} \textbf{\bibinfo{volume}{411}},
  \bibinfo{pages}{L5} (\bibinfo{year}{1993}), \eprint{gr-qc/9304020}.

\bibitem[{\citenamefont{Kokkotas et~al.}(1994)\citenamefont{Kokkotas, Kr\'olak,
  and Tsegas}}]{KKT94}
\bibinfo{author}{\bibfnamefont{K.}~\bibnamefont{Kokkotas}},
  \bibinfo{author}{\bibfnamefont{A.}~\bibnamefont{Kr\'olak}}, \bibnamefont{and}
  \bibinfo{author}{\bibfnamefont{G.}~\bibnamefont{Tsegas}},
  \bibinfo{journal}{Class. Quantum. Grav} \textbf{\bibinfo{volume}{11}},
  \bibinfo{pages}{1901} (\bibinfo{year}{1994}).

\bibitem[{\citenamefont{Kr{\'o}lak et~al.}(1995)\citenamefont{Kr{\'o}lak,
  Kokkotas, and Sch\"afer}}]{KKS95}
\bibinfo{author}{\bibfnamefont{A.}~\bibnamefont{Kr{\'o}lak}},
  \bibinfo{author}{\bibfnamefont{K.}~\bibnamefont{Kokkotas}}, \bibnamefont{and}
  \bibinfo{author}{\bibfnamefont{G.}~\bibnamefont{Sch\"afer}},
  \bibinfo{journal}{Phys. Rev. D} \textbf{\bibinfo{volume}{52}},
  \bibinfo{pages}{2089} (\bibinfo{year}{1995}).

\bibitem[{\citenamefont{Poisson and Will}(1995)}]{PW95}
\bibinfo{author}{\bibfnamefont{E.}~\bibnamefont{Poisson}} \bibnamefont{and}
  \bibinfo{author}{\bibfnamefont{C.}~\bibnamefont{Will}},
  \bibinfo{journal}{Phys. Rev. D} \textbf{\bibinfo{volume}{52}},
  \bibinfo{pages}{848} (\bibinfo{year}{1995}).

\bibitem[{\citenamefont{Balasubramanian
  et~al.}(1995)\citenamefont{Balasubramanian, Sathyaprakash, and
  Dhurandhar}}]{BalSatDhu95}
\bibinfo{author}{\bibfnamefont{R.}~\bibnamefont{Balasubramanian}},
  \bibinfo{author}{\bibfnamefont{B.~S.} \bibnamefont{Sathyaprakash}},
  \bibnamefont{and} \bibinfo{author}{\bibfnamefont{S.~V.}
  \bibnamefont{Dhurandhar}}, \bibinfo{journal}{Pramana}
  \textbf{\bibinfo{volume}{45}}, \bibinfo{pages}{L463} (\bibinfo{year}{1995}),
  \eprint{gr-qc/9508025}.

\bibitem[{\citenamefont{Balasubramanian
  et~al.}(1996)\citenamefont{Balasubramanian, Sathyaprakash, and
  Dhurandhar}}]{BalSatDhu96}
\bibinfo{author}{\bibfnamefont{R.}~\bibnamefont{Balasubramanian}},
  \bibinfo{author}{\bibfnamefont{B.~S.} \bibnamefont{Sathyaprakash}},
  \bibnamefont{and} \bibinfo{author}{\bibfnamefont{S.~V.}
  \bibnamefont{Dhurandhar}}, \bibinfo{journal}{Phys.~Rev.~D}
  \textbf{\bibinfo{volume}{53}}, \bibinfo{pages}{3033} (\bibinfo{year}{1996}),
  \bibinfo{note}{erratum-ibid.~D {\bf 54}, 1860 (1996)},
  \eprint{gr-qc/9508011}.

\bibitem[{\citenamefont{Flanagan and Hughes}(1998)}]{FlanHugh97}
\bibinfo{author}{\bibfnamefont{E.~E.} \bibnamefont{Flanagan}} \bibnamefont{and}
  \bibinfo{author}{\bibfnamefont{S.~A.} \bibnamefont{Hughes}},
  \bibinfo{journal}{Phys. Rev.} \textbf{\bibinfo{volume}{D57}},
  \bibinfo{pages}{4566} (\bibinfo{year}{1998}), \eprint{gr-qc/9710129}.

\bibitem[{\citenamefont{Arun et~al.}(2005)\citenamefont{Arun, Iyer,
  Sathyaprakash, and Sundararajan}}]{AISS05}
\bibinfo{author}{\bibfnamefont{K.~G.} \bibnamefont{Arun}},
  \bibinfo{author}{\bibfnamefont{B.~R.} \bibnamefont{Iyer}},
  \bibinfo{author}{\bibfnamefont{B.~S.} \bibnamefont{Sathyaprakash}},
  \bibnamefont{and} \bibinfo{author}{\bibfnamefont{P.~A.}
  \bibnamefont{Sundararajan}}, \bibinfo{journal}{Phys.~Rev.~D}
  \textbf{\bibinfo{volume}{71}}, \bibinfo{pages}{084008}
  (\bibinfo{year}{2005}), \bibinfo{note}{erratum-ibid. ~{\bf D } 72, 069903
  (2005)}, \eprint{gr-qc/0411146}.

\bibitem[{\citenamefont{Sathyaprakash}(1994)}]{SathyaFilter94}
\bibinfo{author}{\bibfnamefont{B.~S.} \bibnamefont{Sathyaprakash}},
  \bibinfo{journal}{Phys. Rev. D} \textbf{\bibinfo{volume}{50}},
  \bibinfo{pages}{R7111} (\bibinfo{year}{1994}).

\bibitem[{\citenamefont{Owen}(1996)}]{Owen96}
\bibinfo{author}{\bibfnamefont{B.}~\bibnamefont{Owen}}, \bibinfo{journal}{Phys.
  Rev.} \textbf{\bibinfo{volume}{D {\bf 53}}}, \bibinfo{pages}{6749}
  (\bibinfo{year}{1996}).

\bibitem[{\citenamefont{Owen and Sathyaprakash}(1998)}]{OwenSathyaprakash98}
\bibinfo{author}{\bibfnamefont{B.}~\bibnamefont{Owen}} \bibnamefont{and}
  \bibinfo{author}{\bibfnamefont{B.~S.} \bibnamefont{Sathyaprakash}},
  \bibinfo{journal}{Phys.~Rev.} \textbf{\bibinfo{volume}{D \textbf{60}}},
  \bibinfo{pages}{022002} (\bibinfo{year}{1998}).

\bibitem[{\citenamefont{Helstr\"om}(1968)}]{Helstrom68}
\bibinfo{author}{\bibfnamefont{C.}~\bibnamefont{Helstr\"om}},
  \emph{\bibinfo{title}{Statistical Theory of Signal Detection}},
  vol.~\bibinfo{volume}{9} of \emph{\bibinfo{series}{International Series of
  Monographs in Electronics and Instrumentation}} (\bibinfo{publisher}{Pergamon
  Press}, \bibinfo{address}{Oxford, U.K., New York, U.S.A.},
  \bibinfo{year}{1968}), \bibinfo{edition}{2nd} ed.

\bibitem[{\citenamefont{Thorne}(1987)}]{Th300}
\bibinfo{author}{\bibfnamefont{K.~S.} \bibnamefont{Thorne}}, in
  \emph{\bibinfo{booktitle}{Three hundred years of gravitation}}, edited by
  \bibinfo{editor}{\bibfnamefont{S.}~\bibnamefont{Hawking}} \bibnamefont{and}
  \bibinfo{editor}{\bibfnamefont{W.}~\bibnamefont{Israel}}
  (\bibinfo{publisher}{Cambridge University Press}, \bibinfo{year}{1987}), pp.
  \bibinfo{pages}{330--458}.

\bibitem[{\citenamefont{Schutz}(1989)}]{Schutz89}
\bibinfo{author}{\bibfnamefont{B.}~\bibnamefont{Schutz}}, in
  \emph{\bibinfo{booktitle}{The detection of gravitational waves}}, edited by
  \bibinfo{editor}{\bibfnamefont{D.}~\bibnamefont{Blair}}
  (\bibinfo{publisher}{Cambridge University Press}, \bibinfo{address}{England},
  \bibinfo{year}{1989}).

\bibitem[{\citenamefont{Pan et~al.}(2004)\citenamefont{Pan, Buonanno, Chen, and
  Vallisneri}}]{pBCV}
\bibinfo{author}{\bibfnamefont{Y.}~\bibnamefont{Pan}},
  \bibinfo{author}{\bibfnamefont{A.}~\bibnamefont{Buonanno}},
  \bibinfo{author}{\bibfnamefont{Y.}~\bibnamefont{Chen}}, \bibnamefont{and}
  \bibinfo{author}{\bibfnamefont{M.}~\bibnamefont{Vallisneri}},
  \bibinfo{journal}{Phys. Rev. D} \textbf{\bibinfo{volume}{69}},
  \bibinfo{pages}{104017} (\bibinfo{year}{2004}).

\bibitem[{\citenamefont{Compact Binary Coalescence~Group}(2008)}]{CBC}
\bibinfo{author}{\bibfnamefont{L.~S.~C.} \bibnamefont{Compact Binary
  Coalescence~Group}} (\bibinfo{year}{2008}), \bibinfo{note}{private
  Communication}.

\bibitem[{\citenamefont{Perram and Wertheim}(1985)}]{Perram}
\bibinfo{author}{\bibfnamefont{J.}~\bibnamefont{Perram}} \bibnamefont{and}
  \bibinfo{author}{\bibfnamefont{M.}~\bibnamefont{Wertheim}},
  \bibinfo{journal}{J. Comp. Phys.} \textbf{\bibinfo{volume}{{\bf 58}}},
  \bibinfo{pages}{409} (\bibinfo{year}{1985}).

\bibitem[{\citenamefont{{Wikipedia Article}}()}]{WikiBrent}
\bibinfo{author}{\bibnamefont{{Wikipedia Article}}},
  \emph{\bibinfo{title}{{Brent's Method}}},
  \bibinfo{howpublished}{\url{http://en.wikipedia.org/wiki/Brent's_method}}.

\bibitem[{\citenamefont{Brent}(1973)}]{BookBrent}
\bibinfo{author}{\bibfnamefont{R.}~\bibnamefont{Brent}},
  \emph{\bibinfo{title}{Algorithms for Minimization without Derivatives}}
  (\bibinfo{publisher}{Prentice-Hall, Englewood Cliffs, NJ.},
  \bibinfo{year}{1973}), chap.~\bibinfo{chapter}{4}.

\bibitem[{\citenamefont{GSL}()}]{GSL}
\bibinfo{author}{\bibnamefont{GSL}}, \emph{\bibinfo{title}{{The GNU Scientific
  Library is a free numerical library licensed under the GNU GPL}}},
  \bibinfo{howpublished}{\url{http://www.gnu.org/software/gsl/}}.

\bibitem[{\citenamefont{Blanchet
  et~al.}(1995{\natexlab{a}})\citenamefont{Blanchet, Damour, and Iyer}}]{BDI95}
\bibinfo{author}{\bibfnamefont{L.}~\bibnamefont{Blanchet}},
  \bibinfo{author}{\bibfnamefont{T.}~\bibnamefont{Damour}}, \bibnamefont{and}
  \bibinfo{author}{\bibfnamefont{B.~R.} \bibnamefont{Iyer}},
  \bibinfo{journal}{Phys. Rev. D} \textbf{\bibinfo{volume}{51}},
  \bibinfo{pages}{5360} (\bibinfo{year}{1995}{\natexlab{a}}),
  \eprint{gr-qc/9501029}.

\bibitem[{\citenamefont{Blanchet
  et~al.}(1995{\natexlab{b}})\citenamefont{Blanchet, Damour, Iyer, Will, and
  Wiseman}}]{BDIWW95}
\bibinfo{author}{\bibfnamefont{L.}~\bibnamefont{Blanchet}},
  \bibinfo{author}{\bibfnamefont{T.}~\bibnamefont{Damour}},
  \bibinfo{author}{\bibfnamefont{B.~R.} \bibnamefont{Iyer}},
  \bibinfo{author}{\bibfnamefont{C.~M.} \bibnamefont{Will}}, \bibnamefont{and}
  \bibinfo{author}{\bibfnamefont{A.~G.} \bibnamefont{Wiseman}},
  \bibinfo{journal}{Phys. Rev. Lett.} \textbf{\bibinfo{volume}{74}},
  \bibinfo{pages}{3515} (\bibinfo{year}{1995}{\natexlab{b}}),
  \eprint{gr-qc/9501027}.

\bibitem[{\citenamefont{Babak et~al.}(2006)\citenamefont{Babak,
  Balasubramanian, Churches, Cokelaer, and Sathyaprakash}}]{Bank06}
\bibinfo{author}{\bibfnamefont{S.}~\bibnamefont{Babak}},
  \bibinfo{author}{\bibfnamefont{R.}~\bibnamefont{Balasubramanian}},
  \bibinfo{author}{\bibfnamefont{D.}~\bibnamefont{Churches}},
  \bibinfo{author}{\bibfnamefont{T.}~\bibnamefont{Cokelaer}}, \bibnamefont{and}
  \bibinfo{author}{\bibfnamefont{B.}~\bibnamefont{Sathyaprakash}},
  \bibinfo{journal}{Class. Quant. Grav.} \textbf{\bibinfo{volume}{{\bf 23}}},
  \bibinfo{pages}{5477} (\bibinfo{year}{2006}), \eprint{gr-qc/0604037}.

\bibitem[{\citenamefont{Cokelaer}(2007)}]{cokelaer:102004}
\bibinfo{author}{\bibfnamefont{T.}~\bibnamefont{Cokelaer}},
  \bibinfo{journal}{Physical Review D} \textbf{\bibinfo{volume}{76}},
  \bibinfo{eid}{102004} (\bibinfo{year}{2007}).

\bibitem[{LAL()}]{LAL}
\emph{\bibinfo{title}{{LSC Algorithm Library}}},
  \bibinfo{howpublished}{\url{http://www.lsc-group.phys.uwm.edu/daswg/projects%
/lal.html}}.

\end{thebibliography}
\end{document}